\documentclass[graybox,envcountchap, titlepage]{svmult}
\usepackage[square, sort&compress, numbers]{natbib}

\usepackage[pdfpagelabels=true, plainpages=false,
            colorlinks=true, allcolors=blue]{hyperref}

\usepackage{url}
\usepackage{mathptmx}       
\usepackage{helvet}         
\usepackage{courier}        
\usepackage{type1cm}        
\usepackage{makeidx}         
\usepackage{graphicx}        
\usepackage[bottom]{footmisc}

\RequirePackage{doi}
\usepackage{hyperref}
\usepackage{doi}
\usepackage{floatrow}
\usepackage{ragged2e}
\usepackage[rightcaption, ragged]{sidecap}
\usepackage{upgreek}
\usepackage{gensymb}
\usepackage{amsmath}
\usepackage{amssymb}
\usepackage{eucal} 
\usepackage{textcomp}
\usepackage{xcolor}
\newcommand{\hbarhorizline}{\mathchar'26\mkern-7mu h}

\makeindex             

\begin{document}

\title*{Magnetic droplet solitons}
\titlerunning{Magnetic droplet solitons} 

\author{Martina Ahlberg, Sheng Jiang, Roman Khymyn, Sunjae Chung, and Johan~{\AA}kerman}
\authorrunning{Martina Ahlberg \textit{\emph{et al.}}} 
\institute{Martina Ahlberg and Roman Khymyn \at Department of Physics, University of Gothenburg, 41296, Gothenburg, Sweden
\\
Sheng Jiang \at School of Microelectronics, South China University of Technology, 511442 Guangzhou, Chinas
\\
Sunjae Chung \at Department of Physics Education, Korea National University of Education, Cheongju 28173, Korea
\\
Johan {\AA}kerman* \at Department of Physics, University of Gothenburg, 41296, Gothenburg, Sweden, and also with Research Institute of Electrical Communication (RIEC) and Center for Science and Innovation in Spintronics (CSIS), Tohoku University, 2-1-1 Katahira, Aoba-ku, Sendai 980-8577 Japan.
\\(*e-mail: johan.akerman@physics.gu.se)\\}

%
%
\maketitle

\abstract{Magnetic droplets are nanoscale, non-topological, dynamical solitons that can be nucleated in different spintronic devices, such as spin torque nano-oscillators (STNOs) and spin Hall nano-oscillators (SHNOs). This chapter first briefly discusses the theory of spin current driven dissipative magnetic droplets in ferromagnetic thin films with uniaxial anisotropy. We then thoroughly review the research literature on magnetic droplets and their salient features, as measured using electrical, microwave, and synchrotron techniques, and as envisaged by micromagnetic simulations. We also touch upon a closely related soliton, the dynamical skyrmion. Finally, we present an outlook of new routes in droplet science.
\newline
\newline
This is a preprint of the following chapter: Martina Ahlberg, Sheng Jiang, Roman Khymyn, Sunjae Chung, Johan Åkerman, Magnetic Droplet Solitons, published in Nanomagnets as Dynamical Systems, edited by Supriyo Bandyopadhyay, Anjan Barman, 2024, Springer, reproduced with permission of the Springer Nature Switzerland AG 2024. The final authenticated version is available at: \url{https://doi.org/10.1007/978-3-031-73191-4}. }


\tableofcontents

\section{Introduction}
\label{sec:I}

\begin{figure}[htb]
  \begin{center}
  \includegraphics[width=1\textwidth]{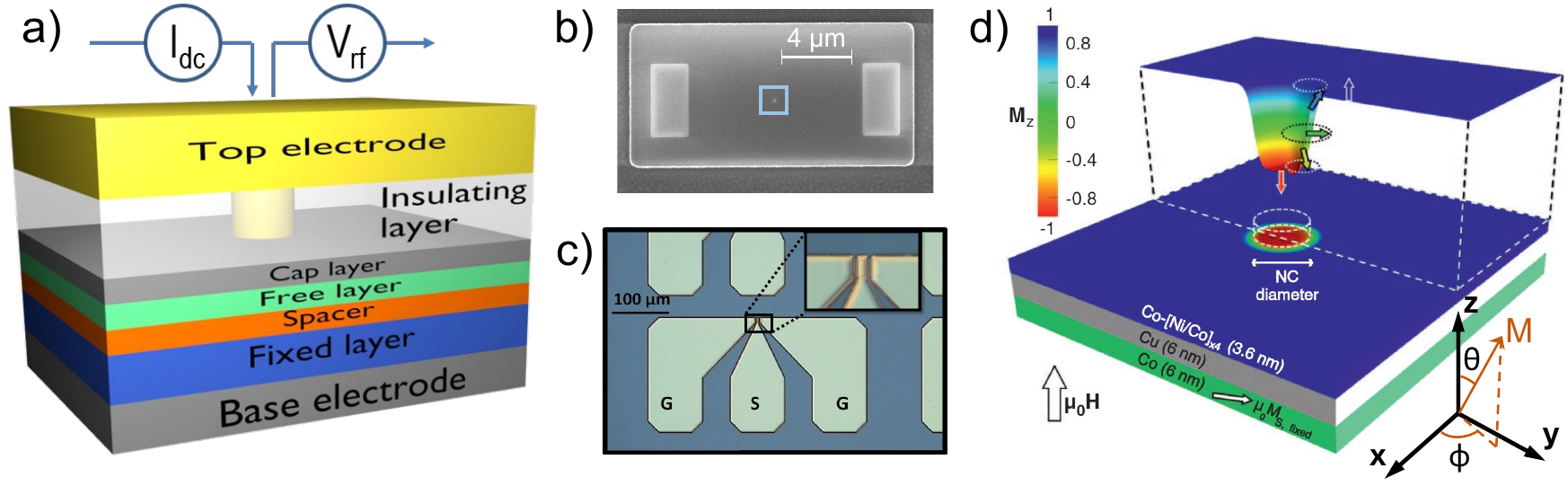}
    \caption{\textbf{a} Schematics of a spin-torque nano-oscillator (STNO). The nanocontact (NC) confines the applied electrical direct current, which becomes spin polarized by the fixed reference layer and, through the resulting spin transfer torque, excites dynamics in the free layer. The free layer magnetodynamics is detected as an rf voltage caused by changes in the device resistance via GMR. \textbf{b} Scanning electron microscope image showing the NC (located within the blue square) and the two ground contacts. The image is taken before the final fabrication step, where the top Au contacts are produced. \textbf{c} Optical microscope image of the Au ground-signal-ground contacts, which are used to connect the device to the measurement setup. The current flows through the NC, down the multilayer stack to the base electrode, where it follows a lateral path until it heads up to the ground contacts. \textbf{d} Illustration of a magnetic droplet. The droplet core magnetization is reversed relative to the surroundings, and the applied out-of-plane field. The spins at the perimeter precess in phase at the characteristic frequency set by the material parameters and the droplet size. \emph{From Mohseni \emph{et al.}}~\cite{Mohseni2013a} \emph{Reprinted with permission from AAAS.}}
    \label{SIFig1}
    \end{center}
\end{figure}

A soliton is a localized and particle-like wave that preserves its shape and is robust against perturbations. Solitons are found in a large variety of systems, but share vital elements such as self-organization through a balance of strong non-linearity and dispersion in the medium.~\cite{Purwins2010} Solitons were first described by John Scott Russell in 1834 as he observed remarkably stable waves in the Union Canal in Scotland, which traveled for miles without losing their shape. The name "soliton" was coined by Zabusky and Kruskal who found localized, self-reinforcing, and highly stable propagating solutions to the Korteweg-de Vries (KdV) equation. Magnetodynamical solitons\footnote{There exists another class of \emph{static} solitons that are \emph{topological}. They are topological particle-like objects in ordered states of matter, which cannot be smoothed out through continuous transformation of the order parameter. In magnetism, they are e.g.~represented by the domain wall, the vortex, the skyrmion, and other topological objects. As our chapter treats magnetodynamical solitons, we will not further discuss static solitons.} 
have been theoretically studied 
for over 40 years, with many early important results 
\cite{Ivanov1989, Kosevich1981, Kosevich1990, Bespyatykh1994}. Conservative solitons (zero damping) in perpendicular anisotropy magnets were analyzed and termed “magnon drops”, in an analogy of a gas compared to a drop of quasi-particles ("free" versus condensed magnons).~\cite{Ivanov1976, Ivanov1977} However, the envisaged phenomenon was never experimentally realized due to the detrimental effect of magnetic damping. Micromagnetic simulations, where the anti-damping effect of spin transfer torque (STT)~\cite{Hoefer2010} was used to cancel the damping locally, suggested that spin torque nano-oscillators (STNOs) should be able to sustain a so-called magnetic droplet. This paved the way for the first experimental demonstration of magnetic droplets, published as late as 2013~\cite{Mohseni2013a}. The distinction “droplet” in contrast to “drop”, was made to emphasize the strongly dissipative nature of the STT-stabilized soliton. The archetype device hosting droplets, the STNO, is outlined in Fig.~\ref{SIFig1}a)--c) together with an illustration of the droplet itself (Fig.~\ref{SIFig1}d)). Since its discovery, this distinctive soliton has continued to attract interest, both from fundamental and application-driven perspectives. While many early aspects of magnetic droplets have been covered by previous reviews on the topic~\cite{Sulymenko2018ltp, Macia2020, Chung2015, Akerman2018}, and can be used as a complement to this chapter, a number of very recently demonstrated novel droplet phenomena are here summarized for the first time.  

In this chapter we give a thorough description of the droplet, starting with the initial prediction, realization, and verification. We continue with the details of generation, inherent dynamics, and means to control and maneuver the droplet. The final subsections portray complementary devices and resemblant solitons. Taken together, the content serves as a valuable introduction for the curious reader, as well as a comprehensible overview of the field.


\section{Pioneer Studies}
\label{sec:PI}

\subsection{Theoretical Prediction}
\label{sec:PI:TP}

Spin wave drops, i.e.~particle-like magnon condensates, were predicted in Refs.~\cite{Ivanov1977,Ivanov1989, Ivanov1976} for the conservative case, i.e.~in the absence of damping. The simplest system where an attraction between magnons can be realized is a uniaxial ferromagnet with easy-axis anisotropy (in the form $K \sin^2\theta$) and an external field $H$ along the anisotropy axis. Thin films with sufficiently strong perpendicular magnetic anisotropy (PMA) to overcome the demagnetizing field are examples of such systems. In this geometry, magnons experience an attractive force and can condense into drops. The attractive interaction emerges due to the uniaxial anisotropy.~\cite{Kosevich1990} The anisotropy term in the Hamiltonian renders a lower energy cost of a new spin deviation (magnon) close to a first magnon, compared to an additional excitation further away.~\cite{Silberglitt1970, Tonegawa1970, Sharma2022} 

We define $\theta$ as the polar angle between the easy-axis and the magnetization, while $\phi$ is the azimuthal angle in the film plane (see Fig.~\ref{SIFig1}d). Throughout the drop, the azimuthal angle of all its spins varies linearly in time as $\phi=\omega t$, with $\omega$ lying between the Zeeman frequency and the ferromagnetic resonance (FMR): $\gamma H < \omega <  \omega_{\text{FMR}}$. The polar angle is both azimuthally symmetric and independent on time $\theta=\theta(r)$, and, hence, at every point, the magnetization precesses with the same phase and with a radially dependent cone angle. The precession frequency and the radial profile are strongly connected and depend on the number of magnons in the condensate. Close to FMR, the soliton profile is quite narrow and the angle at the center $\theta(r=0)=\theta_0$ is small, i.e.~the drop is not yet reversed. However, as more magnons condense, the frequency decreases and $\theta_0$ grows towards the limit value of a reversed core ($\theta_0\rightarrow\pi$), and the drop and its profile $\theta(r)$ expand. Thus, for low droplet frequencies, there is a substantial area at the core of the drop where the magnetization is reversed. The profile of the drop satisfies a balance between exchange and anisotropy. Similar to a domain wall, the exchange length ($\lambda_{\text{ex}}$) defines the width of the transition region between the reversed core and the surroundings aligned with the field.

A dissipative magnetic droplet is closely related to the magnon drop and inherits most of its general properties. A droplet can exist in a magnetic medium \emph{with} damping, as long as there is an energy source available to sustain it. In spin torque nano-oscillators, the droplet is nucleated and sustained by spin transfer torque (STT) provided by a spin-polarized current through a nano-contact. The theory for such magnetic droplets in STNOs was developed in a seminal paper from 2010 by Hoefer, Silva, and Keller~\cite{Hoefer2010}. In addition to the magnon drop's balance between exchange and anisotropy, a droplet also requires a balance between magnetic losses in the form of Gilbert damping and the energy gain provided by the STT. This additional condition selects a certain frequency and thereby the profile of the droplet, as well as a minimum sustaining current. In a fully symmetric system described by rescaled, dimensionless units, the sustaining droplet current is given solely by the nanocontact (NC) radius, apart from a weak influence of the spin torque asymmetry ($\nu$). The NC radius also sets the maximum frequency, the core reversal, and the size of the droplet. Thus, in contrast to the conservative case, where the drop can be of any size, the dissipative droplet is mostly limited by the size of the NC. Smaller droplets exhibit higher frequencies and less core reversal. 

The symmetry is broken by the Oersted field, which shifts the droplet position off-center, leads to spatial magnetic phase variations, and induces a minor frequency reduction. 
A canted applied field, and/or a canted reference layer magnetization, also breaks the symmetry. 
These three perturbations together can trigger the droplet to escape the NC. This so-called \emph{drift instability} is reviewed in Sec.~\ref{sec:D:DI} below. 



\subsection{First Experiments} 
\label{sec:PI:FE}

\begin{figure}[htb]
  \centering
  \includegraphics[width=0.65\textwidth]{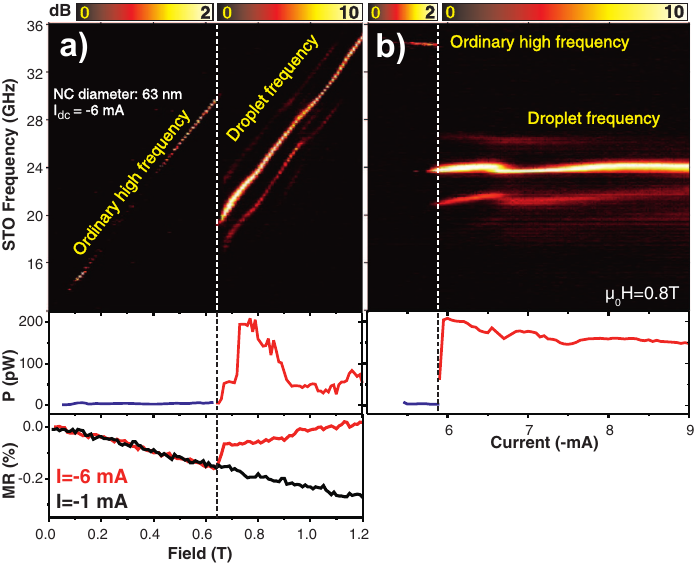}
  \caption{Typical field and current dependence of an orthogonal STNO. \textbf{a} Frequency, power (P), and magnetoresistance (MR) as a function of applied field. The applied current is $-6$~mA, except for the black MR curve, which was recorded at $I=-1$~mA (no droplet). The droplet nucleation is clearly visible at the threshold field $\upmu_{0}H=0.65$~T (marked by the dashed vertical line) as a drop in frequency, a rising P, and a kink in the MR.  \textbf{b} Frequency and power as functions of applied current at a field of $\upmu_{0}H=0.8$~T. The droplet nucleates at the threshold current $I=-5.8$~mA (marked by the dashed vertical line). \emph{From Mohseni \emph{et al.}}~\cite{Mohseni2013a} \emph{Reprinted with permission from AAAS.}}
	\label{SPSFig1}
\end{figure}

Three years after the theoretical prediction, Mohseni \emph{et al.}~reported the first experimental demonstration of a magnetic droplet soliton~\cite{Mohseni2013a}. The authors used orthogonal STNOs, with an out-of-plane (OOP) anisotropy Co/Ni-multilayer as the free layer (FL) and in-plane anisotropy Co as the reference 
layer (RL). The experimental signature of droplet nucleation is remarkably distinct. Figure~\ref{SPSFig1}(a) shows how the power spectral density (PSD), the signal power, and the magnetoresistance (MR) evolve with applied field ($\upmu_{0} H$). Below $\upmu_{0} H = 0.65$~T, a weak high-frequency signal is observed, and the MR decreases linearly as the free and fixed layers become more parallel. Then, at the threshold field, all measured parameters change drastically. The signal frequency drops by 10.3~GHz, the power rises substantially, and the magnetoresistance increases abruptly and its slope reverses sign. Likewise, with increasing current the same type of changes are observed above the critical nucleation current, as illustrated in Fig.~\ref{SPSFig1}(b). These measured characteristics align with the theoretical predictions of the droplet state~\cite{Hoefer2010}. Simply put, this abrupt threshold is the experimental fingerprint of droplet formation.

Another confirmed theoretical prediction was the hysteresis in sustaining current. A lower current is needed to sustain an existing droplet, compared to the initial nucleation current. Thus, there is an energy barrier separating the droplet state from the parallel state. 

The nucleation of a droplet, and its general field dependence, were reproducible from device to device.  However, additional sidebands of the main frequency (Fig.~\ref{SPSFig1} top) were only observed in some devices. Simulations reveal that the sidebands originate from auto-modulation and drift instabilities more sensitive to device-to-device variations, which are topics covered in Sec.~\ref{sec:D:APM} and~\ref{sec:D:DI}.

The authors argued that the nucleation current ($I_{\mathrm{n}}$) should predominantly be driven by the perpendicular component of the polarized spin current, which is directly proportional to the magnitude of the out-of-plane magnetization of the reference layer ($M_{\mathrm{z}}^{\mathrm{RL}}$). The higher the OOP applied field, the higher the $M_{\mathrm{z}}^{\mathrm{RL}}$, and the lower the nucleation current, $I_{\mathrm{n}} \propto 1/ H$. This relation is indeed observed for low fields ($<80\%$ of the reference layer saturation field ($H_{\mathrm{sat}}^{\mathrm{RL}}$)). Later studies have shown that this is only half of the story, see below and Sec.~\ref{sec:GA:NB}.

After the first demonstration, another independent experimental study of magnetic droplets was reported by Maci{\`a} \emph{et al.}~\cite{Macia2014}. This paper explores the droplet evolution using detailed MR measurements at low temperatures (4.2~K). Again, orthogonal STNOs was used, but here permalloy served as the reference layer. The droplet state was recognized by the characteristic abrupt drop in frequency at the nucleation current, and the presence of an energy barrier for droplet formation was verified by current sweeps showing hysteresis. Notably, no hysteresis was found at low fields, and the hysteresis area increased with the applied out-of-plane field. 

\begin{figure}[htb]
  \centering
  \includegraphics[width=1\textwidth]{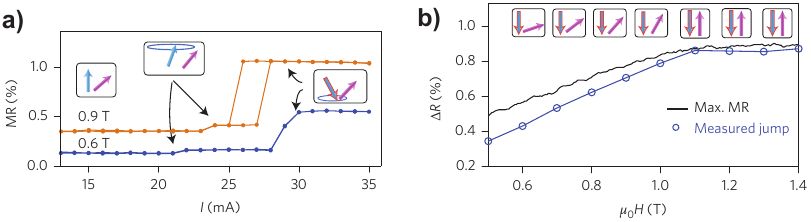}
  \caption{Low temperature ($T = 4.2$~K) magnetoresistance (MR). \textbf{a} The large step in MR corresponds to droplet nucleation, while the tiny increase marks the preceding spin wave excitation. \textbf{b} Comparison of calculated maximum MR and observed steps at droplet nucleation. The droplet core is not fully reversed until the reference layer is aligned to the applied field.~\emph{From Maci{\`{a}} \emph{et al.}}~\cite{Macia2014} \emph{Reprinted with permission from Springer Nature.} }
	\label{SPSFig2}
\end{figure}

The MR showed an excellent signal-to-noise ratio, which allowed the identification of spin wave (SW) excitations, preexisting the droplet at lower currents, see Fig.~\ref{SPSFig2}a. Furthermore, the maximum expected MR (assuming a full free layer reversal beneath the nanocontact) was calculated and compared to the measured step at droplet nucleation. The result is shown in Fig.~\ref{SPSFig2}b. The measured MR is consistently lower than the maximum MR for fields below the reference layer saturating field. As the applied field becomes high enough to saturate the fixed layer out-of-plane, the difference between measured and calculated MR almost vanishes, leaving a 5\% deviation. The result confirms that the droplet core is essentially fully reversed compared to the surrounding magnetization.

This study also examined the field dependence of the nucleation current. Contrary to the findings in Ref.~\cite{Mohseni2013a}, they found a linear trend, which they could fit using a macrospin model. However, the two reports focus on different field ranges: Mohseni \emph{et al.}~\cite{Mohseni2013a} reported on the low field limit, while Maci{\`a} \emph{et al.}~\cite{Macia2014} considered fields above $H_{\mathrm{sat}}^{\mathrm{RL}}$.

\section{Direct Observation}
\label{sec:DO}

The first experimental droplet studies yielded results in good agreement with the theory, which were also corroborated by micromagnetic simulations. However, it was still possible to argue that there were weaknesses in the experimental approach -- all measurements were indirect. The final piece of evidence was missing. To confirm the existence of the dissipative droplet soliton, a direct observation, an image, was needed. 

\begin{figure}
  \begin{center}
  \includegraphics[width=1\textwidth]{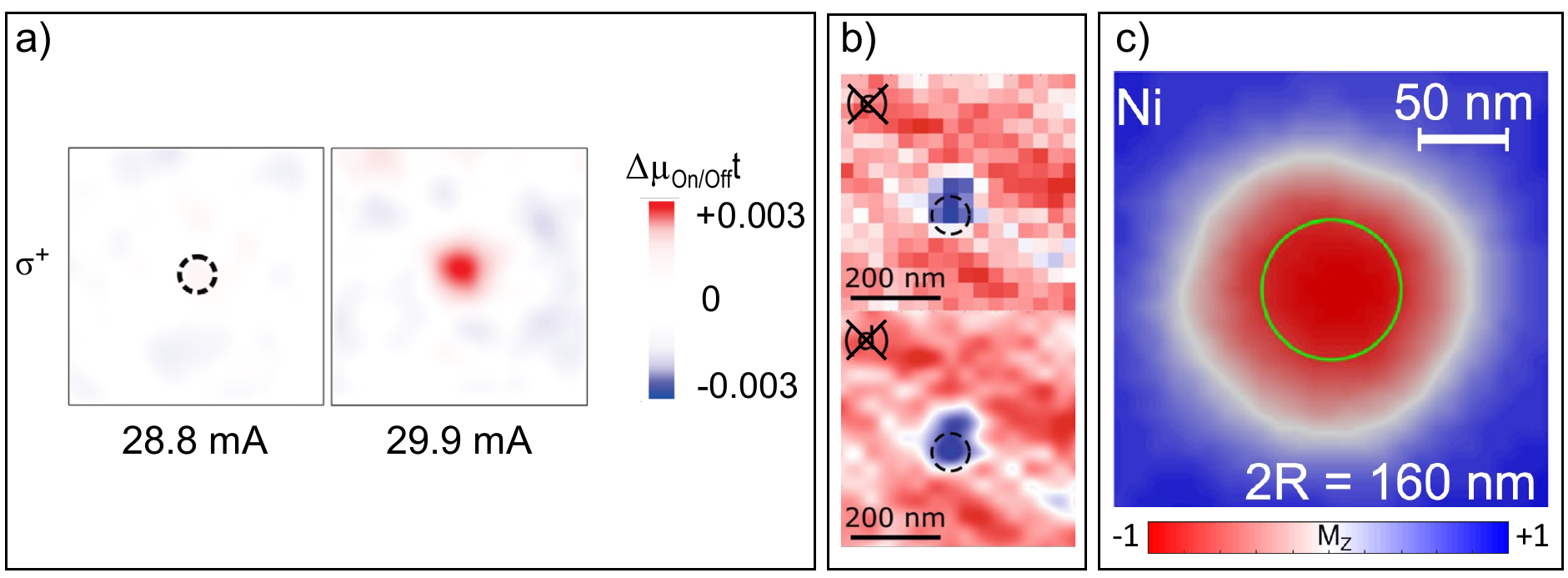}
    \caption{Experimental images of droplets. \textbf{a} STXM contrast below (28.8~mA) and above (29.9~mA) the threshold current. The droplet appears not to be fully reversed, since the image is a time average of a drifting droplet. The dashed circle represents the nanocontact. \emph{Reprinted with permission from Backes \emph{et al.}}~\cite{Backes2015} \emph{Copyright (2015) by the American Physical Society.} \textbf{b} HERALDO image of a droplet. The color scale is missing in the original publication. Time-averaging effects result in a non-circular magnetic structure. \emph{From Burgos-Parra \emph{et al.}} \cite{Burgos2018} \emph{Reprinted under \href{http://creativecommons.org/licenses/by/4.0/}{CC BY 4.0.}} \textbf{c} STXM image of a stable droplet. The large size (twice the nanocontact) is caused by in-plane Zhang-Li torque. \emph{Reprinted with permission from Chung \emph{et al.}}~\cite{Chung2018PRL} \emph{Copyright (2018) by the American Physical Society.}}
    \label{SDOFig1}
    \end{center}
\end{figure}

The droplet radius is only about 100~nm and few imaging techniques are available that provide the required resolution. Scanning transmission X-ray microscopy (STXM) is one of them, and it can also measure very small changes in magnetization, down to $10^{-4}$, as well as an element selective contrast~\cite{Bonetti2015, Macia2012}. Using this method, the first successful direct observation a droplet was made in 2015 by Backes \emph{et al.}~\cite{Backes2015}. They detected a magnetic contrast, which clearly corresponded to a decrease in $M_{\mathrm{z}}^{\mathrm{FL}}$ and only appeared above a threshold current (see Fig.~\ref{SDOFig1}(a)). Furthermore, they were able to confirm that a soliton indeed forms under the nanocontact (NC) by comparing the spin wave excitation profile with calculated predictions. The measured envelope was consistent with a droplet, while propagating SWs could be discarded. As expected, the full-width-half-maximum of the reversed magnetic area matched the NC diameter. Yet, the observed $25\degree$ of reversal was far from the $180\degree$ inferred earlier from magnetoresistance measurements. The origin of this discrepancy lies in the droplet drift instabilities caused by temperature fluctuations and/or a non-saturated fixed layer. During the 60--90 minutes it takes to acquire one image, the droplet may have moved around and even annihilated/re-nucleated. Thus, the final image represents a time average and displays a mean value rather than an instantaneous picture of a fully developed droplet.

Another study used holography combined with extended reference autocorrelation by linear differential operator (HERALDO)~\cite{Guizar2007, Guizar2008} in an attempt to image a droplet~\cite{Burgos2018}. As in STXM, the magnetic contrast is given by X-ray magnetic circular dichroism (XMCD)~\cite{Thole1992, Chen1995}. The paper reports on the evolution of the domain pattern in the $\left[\mathrm{Co/Ni}\right]$ free layer with applied field and current. However, while holography should enable single-shot images, only time-averaged results are presented in the paper, likely to enhance the signal-to-noise ratio. The authors detected a droplet-like feature for a single experimental setting. The image is presented in Fig.~\ref{SDOFig1}(b). The discrepancies between a well-defined circular perimeter and the observed deformed structure were attributed to the time-averaging of the moving droplet. Hence, an experimental image of a stable droplet, which could be compared to simulations, was still lacking in the literature.

Finally, in 2018, the first direct image of a fully reversed magnetic droplet soliton was published (Fig.~\ref{SDOFig1}(c))~\cite{Chung2018PRL}. To the authors' surprise, the droplet was twice as big as the nanocontact. A possible explanation could again be the time-averaging of a droplet moving about the NC. However, theory and simulations predicting a droplet with about the same size as the NC diameter rely on a simplified model of the electrical current distribution within the device, 
commonly described as just a cylinder. However, more realistic COMSOL calculations revealed that the electrical current also has an in-plane radial component. The in-plane component leads to a lateral spin-torque, known as Zhang-Li torque (ZLT), which could account for the droplet enlargement. Further evidence was given by electrical measurements, which showed that the effect was reversed upon reversing the sign of the current and thereby ZLT. In the anti-parallel state, a droplet is nucleated by a positive current and this droplet is instead \emph{compressed} compared to the parallel state droplet. The unveiled importance of the lateral torque inspired a subsequent report on broader effects of ZLT on different modes in spin-torque nano-oscillators.~\cite{Albert2020} 


\section{Generation and Annihilation}
\label{sec:GA}

\subsection{Nucleation Boundaries}
\label{sec:GA:NB}

Droplets form when the applied current generates sufficient spin torque to overcome the magnetic damping in a material with large perpendicular anisotropy. For an all-perpendicular system, ignoring perturbations such as the Oersted field ($H_{\mathrm{Oe}}$), the PMA-to-exchange balance decides the droplet frequency. The nucleation current ($I_{\mathrm{n}}$) is a function of this frequency, which increases linearly with applied field.~\cite{Hoefer2010}. Hence, in the simplest approximation, $I_{\mathrm{n}}$ should only depend on material parameters of the free layer and be directly proportional to $\upmu_{0}H$, as long as the STT overcomes the damping. 

On the other hand, one can also describe droplet nucleation as a consequence of the Slonczewski mode being modulationally unstable, i.e. propagating spin waves collapse into a localized soliton~\cite{Hoefer2010, Chung2016}. The primary requirement is accordingly that the applied current exceeds the SW excitation threshold, which is field dependent~\cite{slonczewski1999jmmm, Slonczewski2002}. This line of argument leads to a similar conclusion as above, but highlights the role of the STT and thereby the reference layer. For all-perpendicular devices, $I_{\mathrm{n}}$ increases linearly with $\upmu_{0}H$, which has also been observed experimentally~\cite{Chung2018PRL}, but for orthogonal devices two different field regimes emerge.

We described in Sec.~\ref{sec:PI:FE} how the two pioneering studies concluded on dissimilar equations for $I_{\mathrm{n}}(H)$. This inconsistency was resolved in a study by Chung~et~al.~\cite{Chung2016}. They followed the work by Slonczewski~\cite{slonczewski1999jmmm} and Hoefer~\cite{Hoefer2010} to derive an equation for the nucleation current:
\begin{equation} 
		\centering
	I_{\mathrm{n}}= \alpha \mathcal{A} +  \frac{\mathcal{B}}{H} + \mathcal{C}
		\label{eq:In}
\end{equation}\

\noindent where $\alpha$ is the damping. All three constants $\mathcal{A}$\footnote{$\mathcal{A} = \frac{\nu j_{0} R^{2}_{\mathrm{NC}}}{M_{\mathrm{s}}}$}, $\mathcal{B}$\footnote{$\mathcal{B} = j_{0} M_{\mathrm{s, p}} \left[ 1.863 l^{2}_{\mathrm{ex}} + \alpha R^{2}_{\mathrm{NC}} \left( \frac{H_{\mathrm{k}}}{M_{\mathrm{s}}} -1 \right) \right]$}, and $\mathcal{C}$\footnote{$\mathcal{C} = j_{0} \left[ \alpha R^{2}_{\mathrm{NC}} \left( \frac{M_{\mathrm{s, p}}}{H} + \nu \left( \frac{H_{\mathrm{k}}}{M_{\mathrm{s}}} -1 \right)\right) + \nu 1.863 l^{2}_{\mathrm{ex}} \right]$} contain material parameters, the NC area, and spin torque coefficients\footnote{$M_{\mathrm{s}}$ is the FL magnetization, $M_{\mathrm{s, p}}$ is the polarizing RL magnetization, $R_{\mathrm{NC}}$ the NC radius, $H_{\mathrm{k}}$ the FL anisotropy field, $l_{\mathrm{ex}}$ the FL exchange length,  $\nu = \left(\lambda^{2} - 1\right)\left(\lambda^{2} + 1 \right)$, and $j_{0} = \left( \left(\lambda^{2} - 1\right)M^{2}_{\mathrm{s}} \mathrm{e} \upmu_{0} \uppi \delta \right) / \left(  \mathrm{ \hbarhorizline} \epsilon \lambda^{2} \right)$, where $\delta$ is the FL thickness, $\mathrm{e}$ is the elementary charge, $\upmu_{0}$ the permeability of free space, and $ \mathrm{ \hbarhorizline }$ the reduced Planck's constant.}.
Specifically, $\mathcal{A}$ and $\mathcal{B}$ include the spin torque asymmetry ($\lambda$) and efficiency ($\epsilon$). Equation~\ref{eq:In} unites the low and high field regions, taking into account both spin wave stiffening ($\propto H$) and reference layer saturation ($\propto 1/H$). 
 
\begin{figure}
  \begin{center}
  \includegraphics[width=1\textwidth]{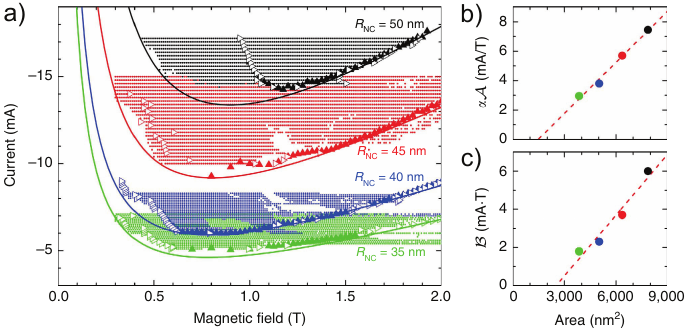}
    \caption{\textbf{a} Droplet nucleation boundaries. The open triangles are determined by field sweeps, the filled triangles by current sweeps, and the small, filled circles correspond to regions displaying low-frequency noise. The different colors represent different NC radii ($R_{\mathrm{NC}}$). The solid lines are fits of Eq.~\ref{eq:In}. The most significant results of the fits are the coefficients \textbf{b} $\alpha \mathcal{A}$ and \textbf{c} $\mathcal{B}$, from which the STT asymmetry $\lambda$ and STT efficiency $\epsilon$ can be obtained by linear fits, represented by the red dashed lines. The x-axis of \textbf{b} and \textbf{c} denotes the NC area. \emph{From Chung \emph{et al.}}~\cite{Chung2016} \emph{Reprinted under \href{http://creativecommons.org/licenses/by/4.0/}{CC BY 4.0.}}}
    \label{SGAFig1}
    \end{center}
\end{figure}

Moreover, their report includes a large compilation of nucleation field/current values, using four different NC radii. The result is found in Fig.~\ref{SGAFig1}(a) together with fits to Eq.~\ref{eq:In}. The dotted areas illustrate regions where the low-frequency noise exceeds 0.3~pW. Figures~\ref{SGAFig1}(b) and (c) display the determined constants $\alpha \mathcal{A}$ and $\mathcal{B}$, which subsequently were used to extract $\lambda=2.38 \pm 0.43$ and $\epsilon = 0.28 \pm 0.05$ of the Co reference layer. Droplet nucleation boundaries thus offer a straightforward method to determine the STT asymmetry.  Furthermore, these values should constitute a better estimate compared to values extracted from single device data. The advantage is that a set of devices is used together to reach the final estimate, compensating random sample-to-sample imperfections. 

The model was later revised by the same research group to include the impact of the Oersted field~\cite{Jiang2018ieee}, which affects the response of the RL magnetization to the applied field. Using calculations where an Oersted-term was added to the equation, they showed that $M_{\text{z}}^{\text{RL}}$ can not be taken as linear in $H$, calling for a modification of Eq.~\ref{eq:In}, which they provided. The extracted values of the revised fit were about $10\%$ lower compared to the original work. The improved model has been used to determine the STT asymmetry and efficiency of $\mathrm{Co}_{x}\mathrm{NiFe}_{1-x}$ thin films~\cite{Jiang2018pra}.

In short, Eq.~\ref{eq:In} describes the droplet nucleation boundaries well, but the Oersted field should be taken into account. It should be noted that the original model also is used in Ref.~\cite{Lendinez2017}, but gives a poor fit, which was resolved by adding a constant offset to the field ($H-H'$). Perhaps the fit could have been improved by carefully calculating $M_{\text{z}}^{\text{RL}}(H)$.

Another aspect of droplet nucleation is the role of the reference layer canting. In orthogonal STNOs, the magnetization angle of the RL is set by the strength of the applied OOP field. Ref.~\cite{Mohseni2018nanotech} compares droplets at a single field and different RL saturation magnetization, resulting in different canting angles. They found that the nucleation time is shorter with a tilted reference layer, while a fully OOP RL gives lower nucleation current, higher droplet precession angles, and a more robust droplet frequency in relation to $I$ and $\upmu_{0}H$. 

A similar question is the role of the applied field canting on systems with an in-plane reference layer. Measurements supported by simulations reveal a critical angle of $\sim 50\degree$ separating spin wave emission at lower angles and droplet nucleation at higher angles. A qualitative explanation of the transition from the propagating to the soliton mode is given by the nonlinear frequency-shift parameter $N$, which changes sign to negative when the field points sufficiently out-of-plane.~\cite{Mohseni2018prb}



\subsection{Hysteresis}
\label{sec:GA:H}

\begin{figure}
  \begin{center}
  \includegraphics[width=1.0\textwidth]{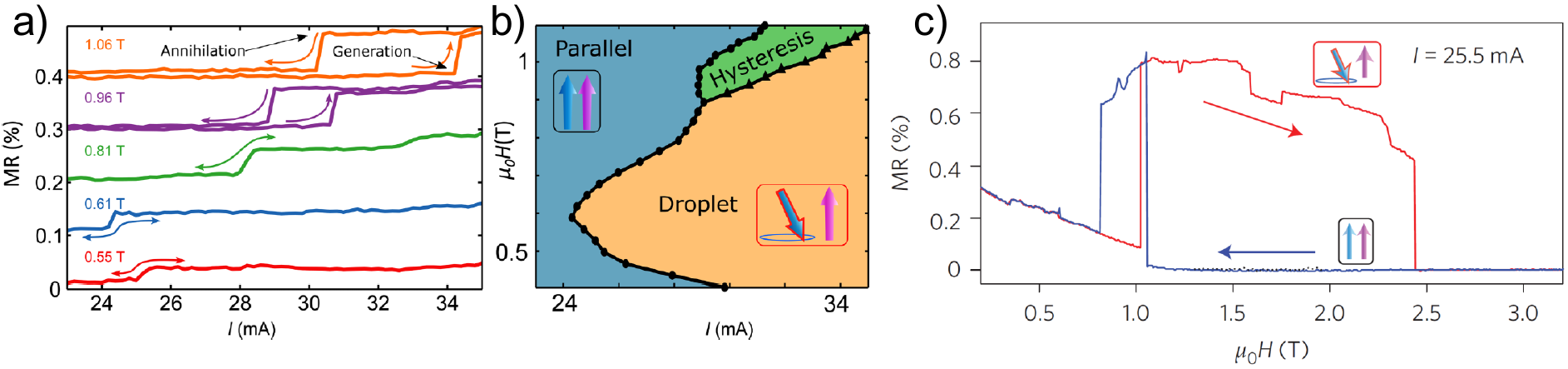}
    \caption{Examples of droplet hysteresis. The measurements are done on orthogonal STNOs. \textbf{a} Current sweeps at different fields marked by individual colors. At the highest fields, the droplet-sustaining current is lower than the nucleation current. \textbf{b} Droplet phase diagram. The blue region denotes the parallel state. The orange and green regions denote the presence of a droplet without and with hysteresis, respectively. \emph{Reprinted figures with permission from S. Lendínez \emph{et al.}}~\cite{Lendinez2015} \emph{Copyright (2015) by the American Physical Society.} \textbf{c} Field scan in a current of $-25.5$~mA. The red line represents the initial increasing field sweep, and the blue line the subsequent decreasing field sweep. The droplet first nucleates at a field of $\upmu_{0} H = 1.02$~T, and annihilates at $\upmu_{0} H = 2.44$~T. When the field decreases the droplet re-nucleates at $\upmu_{0} H = 1.06$~T, and does not collapse again until $\upmu_{0} H = 0.82$~T, below the initial threshold. \emph{From Maci{\`{a}} \emph{et al.}}~\cite{Macia2014} \emph{Reprinted with permission from Springer Nature.}}
    \label{SGAFig2}
    \end{center}
\end{figure}

In the previous section (Sec.~\ref{sec:GA:NB}) we discussed the conditions for droplet nucleation, but these threshold values are not equal to those required to sustain the soliton state, which are lower. Consequently, the droplet is bistable, and the phase diagram shows hysteresis. Already in the first publications, it was observed that the droplet survives in currents below $I_{\mathrm{n}}$~\cite{Hoefer2010, Mohseni2013a, Macia2014}. Figure~\ref{SGAFig2}(a) presents five examples of current sweeps, and Fig.~\ref{SGAFig2}(b) displays a compilation of several measurements. At high fields, there is clear hysteresis, which disappears at moderate fields due to drift instabilities.  

Similarly, at high enough fields, the droplet annihilates but re-nucleates if the field is reduced, and sustains at fields below the initial threshold~\cite{Macia2014}, see Fig.~\ref{SGAFig2}(c). The measurement history is, therefore, important, and one cannot simply exchange the current and field axis of a droplet stability map. 

The inherent annihilation field of a stable droplet is rarely seen in measurements since the droplet experiences drift in a wide range of parameters, especially in orthogonal STNOs. This is exemplified in Figure~2(a) in Ref~\cite{Chung2016}. Out of thirty-one field sweeps, only two deviate from the general trend. These two correspond to a stable state, where the droplet is sustained above the threshold and collapses due to the applied field, not by escaping the NC.
  
The hysteretic region has been used to experimentally investigate the nucleation/annihilation process and the associated time scales~\cite{Hang2018}. The sample was biased with a current corresponding to the bistable state, and the response to current pulses was investigated. The results showed that much longer and stronger pulses are needed to generate a droplet than to destroy it. The timescales differ by orders of magnitude. A larger NC radius increases the nucleation time even further while leaving the annihilation time unchanged. 

Micromagnetic simulations were used to examine the origin of this discrepancy, which was found to be an incubation time for the droplet state~\cite{Hang2018}. After the pulse is applied, nothing happens until 21~ns later whereafter the droplet forms rapidly. The waiting time can be eliminated if the initial magnetization state is destabilized by an in-plane field pulse. The nucleation then occurs at the same time scale as annihilation, a few nanoseconds. Accordingly, the intrinsic times to reconfigure the magnetic state from uniform to droplet or vice versa are the same. Furthermore, adequate spin wave excitations must precede droplet nucleation, and in the absence of external stimuli the excitation only relies on random thermal fluctuations, which implies that the nucleation time increases with decreasing temperature. 

\section{Dynamics}
\label{sec:D}

\subsection{Fundamental Frequency}
\label{sec:D:FF}

The droplet radius and its frequency are intimately related. The larger the droplet, the lower the frequency and the relation can be expressed as~\cite{Xiao2017}:

\begin{equation} 
	\centering
		\omega_{0} = \omega_{H} + \frac{\lambda_{\mathrm{ex}} \omega_{\mathrm{M}}}{\rho_{0}}
		\label{eq:omega0}
\end{equation}\

\noindent where $\omega_{0}$ is the fundamental perimeter frequency, $\omega_{H}=\gamma \upmu_{0} H$,\footnote{In the original publication this relation stated as $\omega_{H}=\gamma H$, where $H$ is incorrectly given in units of T.} $\omega_{\mathrm{M}}=\gamma \upmu_{0} M_{\mathrm{s}}$, $\gamma$ is the gyromagnetic ratio, $H$ the perpendicular applied field,  $\upmu_{0}$ is the permeability of free space, $\rho_{0}$ is the full-width half-maximum radius of the droplet and  $M_{\mathrm{s}}$ is the saturation magnetization, $\lambda_{\mathrm{ex}}=\sqrt{A_{\mathrm{ex}} K}$ is the exchange length, $A_\mathrm{ex}$ is the exchange constant, and $K$ the anisotropy constant of the free layer. The spin precession is largest at the perimeter, while the center of a fully reversed droplet is practically static. 

The equation was first derived for the conservative drop and should be regarded as approximate. It does not account for e.g. long-range magnetostatic interactions nor the strength of the current. Analytics and simulations show that droplets in thicker films display lower frequencies~\cite{Bookman2013, Yazdi2021}. Theoretically, an increasing current is inferred to induce a red shift~\cite{Hoefer2010}, but the experimental data reveal a dispersed behavior. With increasing current the frequency has been observed to decrease~\cite{Mohseni2014, Xiao2017, Mohseni2020prb}, stay virtually constant~\cite{Mohseni2013a, Xiao2017}, vary slightly up and down~\cite{Chung2016, Jiang2018pra}, increase~\cite{Hang2018, Macia2014}, and increase with some discontinuous drops~\cite{Lendinez2015, Statuto2019, Mohseni2013a}. Nonetheless, a fundamental relation is that larger droplets precess with lower frequency~\cite{Hoefer2010}.  


\subsection{Drift instabilities}
\label{sec:D:DI}
\begin{figure}
  \begin{center}
  \includegraphics[width=1\textwidth]{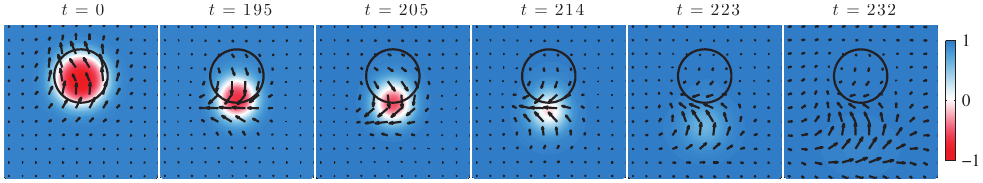}
    \caption{Micromagnetic simulation of a droplet escaping the NC due to a drift instability in a tilted applied field. The frames are captured at different time steps. The droplet dislodges downwards, while the field is directed to the right. Using the same parameters, but a 56\% stronger current results in a stable droplet state. The in-plane vectors are normalized to the maximum value of each frame for clarity. \emph{Reprinted with permission from Hoefer \emph{et al.}}~\cite{Hoefer2010} \emph{Copyright (2010) by the American Physical Society.}}
    \label{SDFig1}
    \end{center}
\end{figure}

A droplet residing in a uniform magnetic environment is very stable, with the in-plane, in-phase precession at the perimeter being the only dynamics~\cite{Hoefer2010}. Nevertheless, real samples are rarely uniform. In addition to potential grain boundaries and distribution of material parameters, the current flow may be non-symmetric, and the ever-present Oersted field modifies the energy landscape. The consequence of inhomogeneities are different types of shape perturbations and dislocations. The notation "drift instabilities" refers to events where the droplet escapes the nanocontact completely. Figure~\ref{SDFig1} displays such an event, where the applied field and fixed layer magnetization are slightly tilted along the $x$-direction, $\theta_{0}=5\degree$ and $\theta_{\mathrm{f}}=31.4\degree$, respectively. The droplet dislodges downwards, i.e. along the $y$-axis. The drift instability disappears at higher currents, leaving a stable droplet state.    

Drift instabilities were originally predicted to arise in certain regions of the available parameter space, although it appeared complex to designate which regions are stable and which are not~\cite{Hoefer2010}. Later theoretical work uncovered well-defined boundaries as functions of current and field, where the droplet is either linearly stable or deterministically prone to drift. The deterministic drift instability regime develops in the region of high current and field. Furthermore, thermal fluctuations give rise to additional rare events where the droplet experiences drift. Factors that promote a stable state are a large NC, low field, moderate current, and high perpendicular anisotropy.~\cite{Wills2016} 

The significance of temperature was further investigated, and the authors could conclude "Consequently, we have shown that the droplet's "drift" instability is unavoidable in a finite-temperature device." But they also stated that rare events can be reduced.~\cite{Moore2019} The significance of the RL magnetization canting angle has not been systematically investigated, although it was included in the earliest paper~\cite{Hoefer2010}. In both Ref.~\cite{Wills2016} and~\cite{Moore2019} the reference layer magnetization was considered to be parallel to the applied field, perpendicular to the sample plane. 

Initial experimental studies only probed the high-frequency dynamics, and overlooked any signals arising from slow drift instabilities~\cite{Mohseni2013a, Macia2014, Mohseni2014, Chung2014}. Albeit drift was acknowledged, it was believed that hysteresis warrants stable droplets. This idea was challenged by experiments and micromagnetic simulations, which proved that low-frequency signals (hundreds of MHz) can arise from periodic drift instabilities even in the hysteretic regime, since the droplet is inclined to re-nucleate as long as there still is some SW precession underneath the nanocontact.~\cite{Lendinez2015}. 

Later low-frequency noise became identified as a characteristic of the droplet~\cite{Chung2016} and a means to distinguish the soliton state in all-perpendicular STNOs~\cite{Chung2018PRL, Ahlberg2022, Shi2022}. However, the droplet does not experience drift in all parameter space, but there exist field and current combinations where the low-frequency power is zero, and the designation of stable/unstable regions is indeed complex~\cite{Jiang2018pra, Ahlberg2022}. Moreover, although the details of the measured noise are very reproducible for an individual device, it differs between samples~\cite{Ahlberg2022}. Further experiments are needed to resolve all aspects of drift instabilities.  


\subsection{Multiple Modes}
\label{sec:D:MM}

Droplets in real materials with defects and disorder, cannot be described simply by a single set of material parameters and a few well-defined perturbations, such as the Oersted field and temperature. On the contrary, reproducible kinks in the magnetoresistance~\cite{Chung2014, Macia2014, Lendinez2015, Chung2016} as well as signals deviating from a continuous function~\cite{Mohseni2013a, Mohseni2014, Lendinez2015}, are commonly found in experiments. These variations are most likely connected to droplet mode transitions, ultimately caused by the complex energy landscape.  

Statuto~\emph{et al.}~\cite{Statuto2019} examined devices displaying a number of separate fundamental mode frequencies, manifested in both field and current sweeps. The authors linked the low-frequency noise to the high-frequency behavior and identified distinct droplet modes. The modes are tied to particularities in the energy landscape.  Moreover, droplets at low currents may be very stable with no concomitant low-frequency noise, not even during mode transitions. In contrast, noise is always present when the droplet experiences instabilities, and then the noise amplitude increases substantially when the droplet shifts from one mode to another. 

Similarly, kinks in the MR usually reflect mode transitions with concomitant low-frequency noise. This has been observed for symmetric all-perpendicular STNOs~\cite{Ahlberg2022}. Traces of mode transitions are clearly seen in the integrated noise map presented in Fig.~\ref{SMFig1}b) in Sec.~\ref{sec:M:F}. Well-defined droplet modes have also been reported in~Ref.~\cite{Mohseni2014}.


\subsection{Automodulation and Perimeter Modes}
\label{sec:D:APM}

\begin{figure}
  \begin{center}
  \includegraphics[width=1\textwidth]{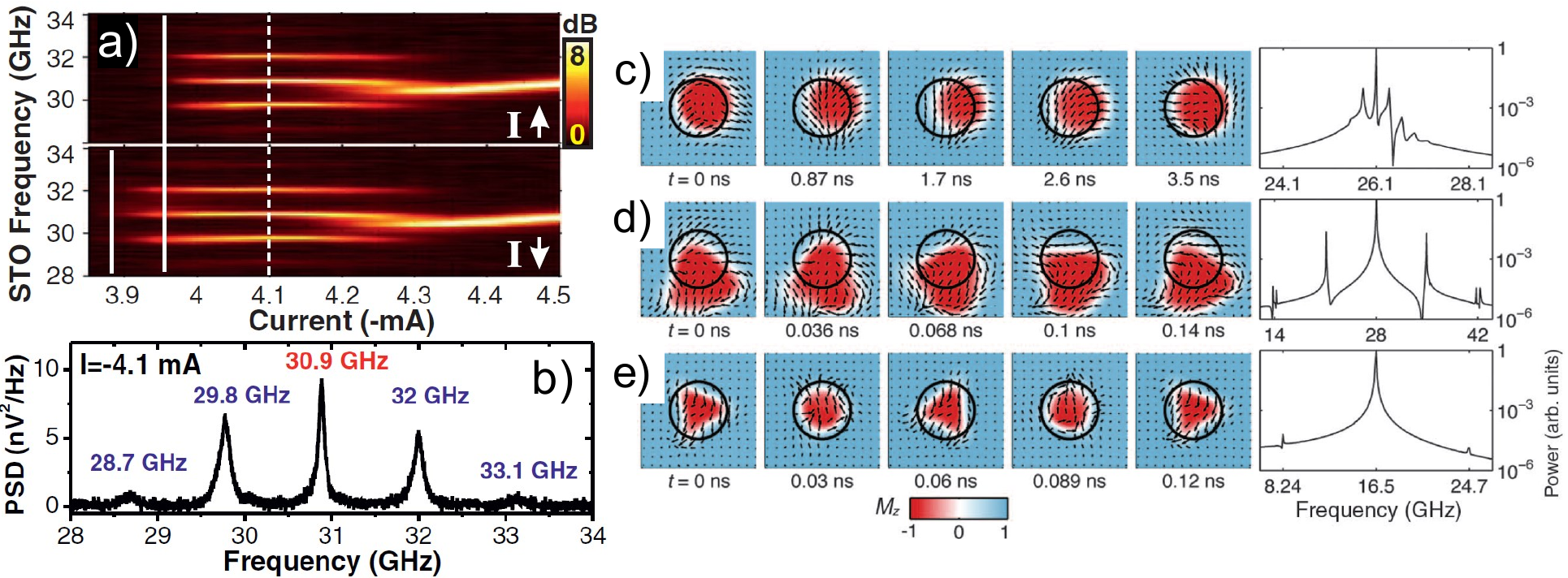}
    \caption{Illustration of sidebands and their origins. \textbf{a} STNO frequency as function of increasing (top) and decreasing (bottom) current. Strong sidebands are present at low current, and the mode shows hysteresis just as a stable droplet. \textbf{b} PSD corresponding to the dashed white line in \textbf{a}. \textbf{c}--\textbf{e} Time sequences of different droplet modes, which produce sidebands: \textbf{c} A gyrating droplet ($\upmu H=0.8$~T, $R_{\text{NC}}=63$~nm, and $I=-8$~mA). \textbf{d} Asymmetric droplet spinning on the NC edge ($0.9$~T, $50$~nm, and $-9$~mA). \textbf{e} Breathing droplet ($0.5$~T, $80$~nm, and $-8$~mA). The right-hand graphs show the Fast Fourier Transform of the magnetization time traces. \emph{From Mohseni \emph{et al.}}~\cite{Mohseni2013a} \emph{Reprinted with permission from AAAS.}}
    \label{SDFig2}
    \end{center}
\end{figure}

The droplet's main frequency is sometimes accompanied by modulation sidebands. They have been observed in both early simulations~\cite{Finocchio2013} and experiments. An example of the latter is found in Fig.~\ref{SDFig2}(a) and the cut along the dashed white line is displayed in Fig.~\ref{SDFig2}(b). Since the external drive is simply a dc current, the sidebands must arise from intrinsic automodulation. Simulations tuned to specific sets of parameters unfolded a number of scenes corresponding to GHz-signals outside the main frequency. Figure~\ref{SDFig2}(c) illustrates how the droplet experiences an approximate circular (gyrotropic) motion underneath the NC. Another scenario is displayed in Fig.~\ref{SDFig2}(d) where the droplet rotates at the NC edge, producing concomitant propagating spin waves. This situation gives rise to clear sidebands. Lastly, Fig.~\ref{SDFig2}(e) reveals a breathing mode associated with frequencies at $1/2$ and $1/3$ of the fundamental mode.~\cite{Mohseni2013a}

A new perspective on the appearance of sidebands was presented in a later publication~\cite{Mohseni2020prb}. The droplet motion was described in terms of inertia and effective mass, giving similar conclusions as numerical studies discussing “the basin of attraction”~\cite{Wills2016, Moore2019}. That is, the droplet experiences a restoring force when field gradients drive it out of the NC area. Both views establish that droplets in large nanocontacts and high perpendicular fields are less susceptible to drift. The simulations in Ref.~\cite{Mohseni2020prb} expose that inertia responding to the Oersted field force leads to additional modes evidenced by sideband signals. Mapping of the spatial profiles of the sideband frequencies unveils that the sideband mode with $f$ above the main frequency precesses clockwise, while the low-$f$ mode precesses counterclockwise. 

\begin{figure}
  \begin{center}
  \includegraphics[width=1\textwidth]{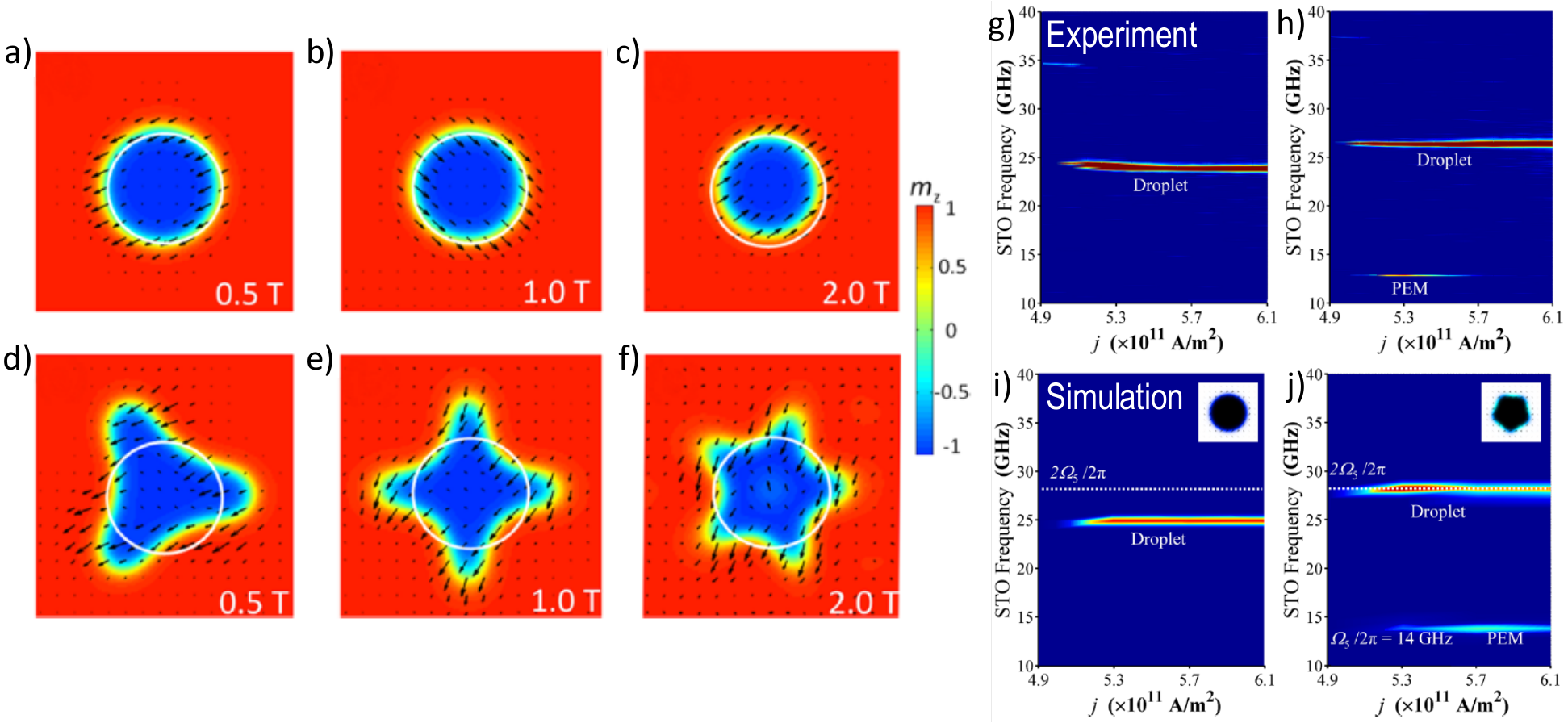}
    \caption{Perimeter excitation mode (PEM) characteristics. \textbf{a}--\textbf{c} Ordinary droplets formed by a low current density ($j = 0.2 \cdot 10^{12}$~A/m$^{2}$), whose shapes are little affected by the external field. \textbf{d}--\textbf{f} PEMs are excited at a higher current density ($j = 2.5 \cdot 10^{12}$~A/m$^{2}$). More perimeter protrusions emerge as the field is increased. \textbf{g}--\textbf{h} Experimental PSD at \textbf{g} $\upmu_{0}H=0.8$~T and \textbf{h} $\upmu_{0}H=0.9$~T with a tilt angle of $7.5\degree$. At the higher field a clear PEM signal is visible beneath the fundamental frequency. \textbf{i}--\textbf{j} Simulated PSD at \textbf{i} $\upmu_{0}H=0.7$~T and \textbf{j} $\upmu_{0}H=0.8$~T with a tilt angle of $3\degree$, together with snapshots of the droplet shape. The PEM is excited when twice its frequency matches the fundamental mode. \emph{Reprinted figures with permission from Xiao \emph{et al.}}~\cite{Xiao2017} \emph{Copyright (2017) by the American Physical Society.}}
    \label{SDFig3}
    \end{center}
\end{figure}

Further dynamics have been experimentally detected. For certain conditions, a single frequency emerges below the fundamental signal. This frequency is associated with perimeter excitation modes (PEMs), that are ordered perimeter distortions. Ordinary stable droplets are depicted in Fig.~\ref{SDFig3}(a)--(c), and PEMs in (d)--(f). To excite these modes the magnetization of the reference layer needs to be nearly perpendicular to the sample plane. That condition fulfilled, a particular PEM will evolve when the fundamental droplet frequency is close to double that of the PEM, which signifies that the process is parametric. Fig.~\ref{SDFig3}(g)--(h) and (i)--(j) represents experimental and simulation data, respectively. The data convincingly corroborate the existence of PEMs, and also show that they disappear at high currents (Fig.~\ref{SDFig3}(h)).~\cite{Xiao2017} Signals associated with PEMs have been measured in Ref.~\cite{Mohseni2020prb} as well, and possibly in Ref.~\cite{Jiang2018pra}. In addition, these modes are readily encountered in simulations.

\section{Control and Maneuvering}
\label{sec:M}

\subsection{Propagation and Collisions}
\label{sec:M:PC}

Propagating and/or interacting droplets has thus far not been studied experimentally, but theoretical work has unveiled intriguing features which warrant attention. 

Droplets can be made to propagate by a field gradient, and the propagation will be in the direction of the lower field.~\cite{Hoefer2012} In the absence of damping a moving droplet is robust and keeps its size, frequency, and velocity even if it is exposed to significant perturbations~\cite{Hoefer2012physD}. However, damping is always present, albeit small and/or effectively tuned to zero by spin currents. The propagation of droplets in realistic settings has been studied analytically together with micromagnetic simulations. Under the action of damping, the droplet velocity accelerates while its size decreases and the frequency increases accordingly. When the frequency reaches the ferromagnetic resonance, the droplet collapses into spin waves. Nonetheless, by using a feed-back controlled alternating applied field, the droplet motion can be stabilized and the velocity kept constant.~\cite{Hoefer2012} Analytical expressions for propagating droplets are also provided in Ref.~\cite{Bookman2015}.

An effective field gradient can also be realized by alteration of the host material’s anisotropy. In Ref.~\cite{Mohseni2020pra} they model a waveguide where the material surrounding the NC has a higher anisotropy than the area next to the NC designed for droplet propagation. The droplet escapes the NC due to the gradient and propagates in an almost straight line only distorted by small wiggles around the main direction. Despite moderate damping the droplet can travel up to a micrometer until it eventually annihilates. The annihilation process results in a burst of spin waves, whose wave-vectors are set by the droplet’s initial velocity. The moving droplet can thus be used as a mobile SW source with a given emission direction. Spin Hall nano-oscillators can also host droplets, which spontaneously leave the active area and propagate for more than 1.5~$\upmu$m.~\cite{Divinskiy2017}

\begin{figure}
  \begin{center}
  \includegraphics[width=1\textwidth]{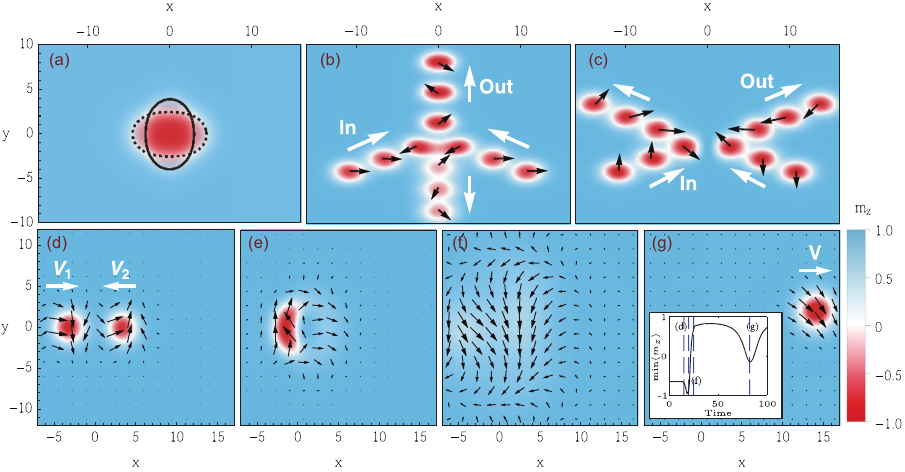}
    \caption{Examples of droplet collision. \textbf{a} Low velocity droplets merge and form a breathing mode (displayed at two moments in time). \textbf{b} Two colliding in-phase droplets temporarily merge, unmerge, and continue in opposite directions along the symmetry axis. \textbf{c} Two droplets with opposite phase repel each other, and reflect along the symmetry axis. \textbf{d}--\textbf{g} The merging process of two droplets with a difference in phase close to the critical value $\Delta_{\mathrm{cr}}$. \textbf{d} The droplets approach each other. \textbf{e} They merge, and \textbf{f} annihilate into spin waves. \textbf{g} A new propagating droplet nucleates in the SW background. \emph{Reprinted with permission from Maiden \emph{et al.}}~\cite{Maiden2014} \emph{Copyright (2014) by the American Physical Society.}}
    \label{SPCFig1}
    \end{center}
\end{figure}

An often-stated attribute of solitons is that they are unaffected by collisions, keeping their original shape. However, this is only true for classical, conservative solitons and should not be assumed valid for dissipative states.~\cite{Purwins2010} 

Collisions between two droplets can be categorized by their relative phase (of the perimeter precession) and momenta (velocities). They will experience an attractive interaction if they are in-phase, and a repulsive interaction if they are out-of-phase. For intermediate phases they are either attracted or repulsed; the two forms of interaction are separated by a critical phase difference $\Delta_{\mathrm{cr}}$. Four typical scenarios have been described. i) Small velocity, small phase difference: the droplets merge and form a breathing mode, see Fig.~\ref{SPCFig1}(a). ii) Large velocity, small phase difference: the droplets merge, unmerge and continue in opposite directions along the symmetry axis, see Fig.~\ref{SPCFig1}(b). iii) Any velocity, the phase difference larger than $\Delta_{\mathrm{cr}}$: reflection reminiscent of Bragg’s law, see Fig.~\ref{SPCFig1}(c). iv) Any velocity, the phase difference close to $\Delta_{\mathrm{cr}}$: the droplets merge, collapse into spin waves, and re-nucleates as a moving single droplet, see Fig.~\ref{SPCFig1}(d--g).~\cite{Maiden2014} 
 

\subsection{Routing}
Droplet propagation can be controlled by utilizing voltage controlled magnetic anisotropy (VCMA). Micromagnetic simulations have been used to outline a routing device for magnetic droplets. The envisioned device consists of a MgO/Fe/MgO trilayer shaped into a small rectangular waveguide connected to a larger routing area. A circular gate electrode and a semiconducting nanostripe are placed on the waveguide and the routing area, respectively. First, the droplet is nucleated underneath the circular electrode by applying a nanosecond voltage pulse, which temporarily reduces the PMA. A small constant in-plane field is also used to further destabilize the OOP magnetization. The droplet is accelerated by the gradient of the demagnetization field in the waveguide and propagates toward the stripe electrode. Next, a semiconducting stripe is biased by a dc voltage giving rise to a PMA gradient perpendicular to the droplet path. The droplet turns by an angle set by the dc bias, changing the direction up to $\pm 70$\degree. The maximum propagation length is $\sim 1~\upmu$m. Consequently, an input voltage can be used to route the droplet to different output MTJ gates where it can be electrically detected.~\cite{Nikitchenko2022} 


\subsection{Merging}
\label{sec:M:M}

Interactions between lateral droplet pairs can also occur if two neighboring nanocontacts host one droplet each. If they are close enough, the droplets may merge. This process is the topic of Ref.~\cite{Xiao2016}. They use zero-temperature simulations of an all-perpendicular system to map out the conditions of merging. The maximum NC spacing is a function of applied field and current. It decreases with field and increases with current, reaching a plateau. The merging process is reversible, but it takes very high fields, up to 7--8~T, to re-separate the droplets. An intermediate state precedes the separation. At medium fields the merged droplet is highly unstable and takes a hollow shape, resembling a donut. Once disconnected, the droplets stay detached while the field is decreased down to $0.2$~T where they again merge. Thus, the process show hysteresis. A notable similarity to the interactions during collisions is that the droplets must precess in-phase to merge. The pair spontaneously synchronize if they are generated simultaneously, but if they form at different times, they may develop opposite phase making merging impossible. External manipulation of the droplet synchronization is possible by means of mutual lock-in to an oscillating microwave magnetic field.~\cite{Wang2017}


\subsection{Freezing}
\label{sec:M:F}

Early droplet research consistently used orthogonal STNOs, since they allow for detection of the fundamental precession frequency. However, this hindered the exploration of the low field limit of the droplet phase diagram. The knowledge about this regime only relied on simulations and they suggested that the droplet should be stable in zero field and only annihilate once the current was switched off.~\cite{Macia2014} Recently, Ahlberg \emph{et al.} have utilized all-perpendicular devices to experimentally access the full field range of droplet existence.~\cite{Ahlberg2022} They discovered that the droplet freezes at low fields into a static nanobubble, which is stabilized by domain wall pinning to magnetic inhomogeneities. Moreover, the nanobubble survives in zero current and persists for days, if not longer. Their conclusions based on in-house electrical measurements were confirmed by STXM images.

\begin{figure}[t]
  \begin{center}
  \includegraphics[width=1\textwidth]{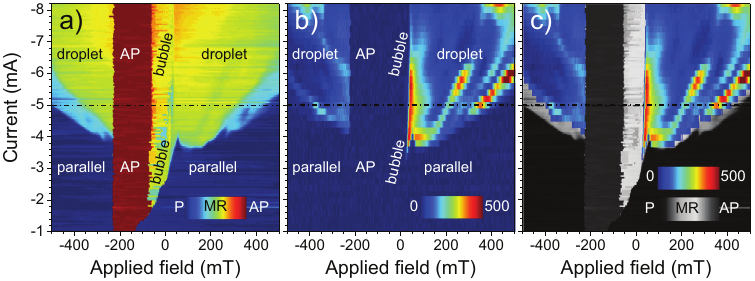}
    \caption{Droplet phase diagram, demonstrating plain parallel and anti-parallel (AP) alignment of the FL and RL, droplet modes, and the static bubble (frozen droplet) state at low fields. The data is compiled of several field scans $0.5 \rightarrow -0.5$~T at constant $I= -1 \rightarrow -8.2$~mA. \textbf{a} Magnetoresistance. Intermediate resistance reveals droplet formation. \textbf{b} Integrated ($0 - 0.5$~GHz) low frequency noise. The bubble state is preceded by erratic dynamics producing strong noise. In addition, traces of large noise are also visible in the droplet regime. These signals are marks of mode transitions.  \textbf{c} The noise map in \textbf{b} overlaid on the magnetoresistance in \textbf{a}. The subtle vertical line in the MR data perfectly corresponds to the disappearance of noise, which signifies the bubble state. \emph{From Ahlberg \emph{et al.}}~\cite{Ahlberg2022} \emph{Reprinted under \href{http://creativecommons.org/licenses/by/4.0/}{CC BY 4.0.}}}
    \label{SMFig1}
    \end{center}
\end{figure}

Figure~\ref{SMFig1} presents a droplet phase diagram based on measurements of the magnetoresistance and integrated low-frequency noise. The low-frequency noise emerges when the droplet moves around the nanocontact, or experiences continuing annihilation/re-nucleation. The nanobubble is static and produces no microwave signal. As seen in the figure (Fig.~\ref{SMFig1}), the bubble nucleates already at currents below the droplet threshold of $\sim 3.8$~mA. At larger currents the droplet state preexists the bubble, and the transition generates large noise. The noise arises from erratic dynamics linked to sporadic pinning to magnetic defects. The transition droplet--bubble is fully reversible without any hysteresis, and thus there is no energy barrier between the two states. Consequently, it is possible to toggle between a dynamic and a static mode, and save the latter, which could be of interest for potential applications.

%
%
%
%
%

\section{Miscellaneous Devices}
\label{sec:MD}

\subsection{Voltage controlled magnetic anisotropy (VCMA)}
\label{sec:MD:VCMA}

One approach to control the magnetization dynamics is to employ voltage-controlled magnetic anisotropy~\cite{Wang2012, Shiota2012, Verba2014}. Simulations show that this effect provides an effective method to tune the auto-oscillation mode from droplet to propagating spin waves. 

The dynamics has been mapped out as a function of electric field $E/E_{0}$, where $E_{0}$ corresponds to zero effective PMA.~\cite{Zheng2020} In contrast to the commonly used layout of an NC in an extended film, the simulated device had a well-defined radius of 256~nm, i.e. a nanopillar structure. This means that the SWs can be reflected at the device boundary and confinement is important.

At a low electrical field, an ordinary droplet is formed, which displays the usual sub-FMR frequency. However, at $E=0.28E_{0}$ the droplet frequency crosses the FMR. Further increase in $E$ leads to a droplet mode with a frequency significantly above FMR. The droplet becomes distorted, its diameter decreases, and the in-plane precession coherency is lost, but the $M_{\mathrm{z}}$ component stays reversed. Thus, the mode can still be regarded as localized, whereas the unexpectedly high frequency must originate from an interplay between the droplet and simultaneously excited spin waves. Above $E=0.45E_{0}$ the localization vanishes and the dynamics consists solely of propagating spin waves.  

\begin{figure}
  \begin{center}
  \includegraphics[width=1\textwidth]{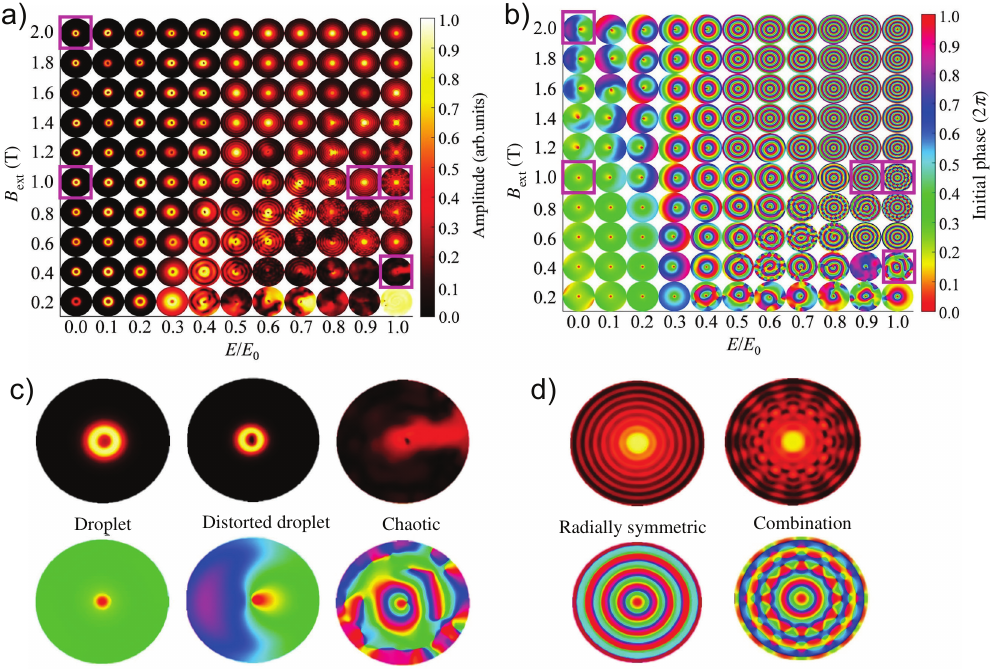}
    \caption{Phase diagram as function of electric ($E/E_{0}$) and magnetic ($B_{\text{ext}}$) field. \textbf{a} Spatial map of the magnetization Fourier amplitude. \textbf{b} Spatial map of the magnetization Fourier phase. The pink squares show examples of \textbf{c} the Droplet (middle, left), Distorted droplet (top, left), and Chaotic SW modes (bottom, right), as well as \textbf{d} a Radially symmetric SW mode (middle, right), and a combination of radially and azimuthally symmetric modes (middle, right). \emph{Reprinted with permission from Zheng \emph{et al.}}~\cite{Zheng2020} \emph{Copyright (2020) by the American Physical Society.}}
    \label{SMDFig2}
    \end{center}
\end{figure}

Figure~\ref{SMDFig2} displays an exhaustive phase diagram of the dynamic state as a function of $E/E_{0}$ and applied field, constructed of Fourier transform images of the excited modes.  Figure~\ref{SMDFig2}(a) show the Fourier amplitude, and Fig.~\ref{SMDFig2}(b) the Fourier phase. Different enlarged examples are found in Fig.~\ref{SMDFig2}(c) and (d). At low magnetic and electric fields, the ordinary droplet is stable. In higher magnetic fields, the main frequency crosses the FMR and a distorted droplet is present. High $E/E_{0}$, i.e. low PMA, facilitates propagating spin waves, but without a strong stabilizing $B$-field, the coherent SWs collapse into chaotic patterns. 

The simulations described above did not take the Oersted field into account, which can be justified by the small magnitude of $H_{\text{Oe}}$ compared to the edge demagnetization field in the nanopillar geometry. A later publication included the Oersted field and discovered a new droplet-like configuration, which was denoted a spiral mode.~\cite{Zheng2021} This mode appears at low currents and is attributed to a competition between $H_{\text{Oe}}$, which is directed clock-wise, and the inertia of the inherent STT-driven droplet precession, which is counter-clock-wise. The spiral mode only exists for quite low $E/E_{0}$ and currents, while the distorted droplet emerges if the PMA is reduced. Both modes transform into an ordinary droplet at high currents.


\subsection{Exchange Spring Magnet}
\label{sec:MD:ExS}

\begin{figure}[b]
  \begin{center}
  \includegraphics[width=1.0\textwidth]{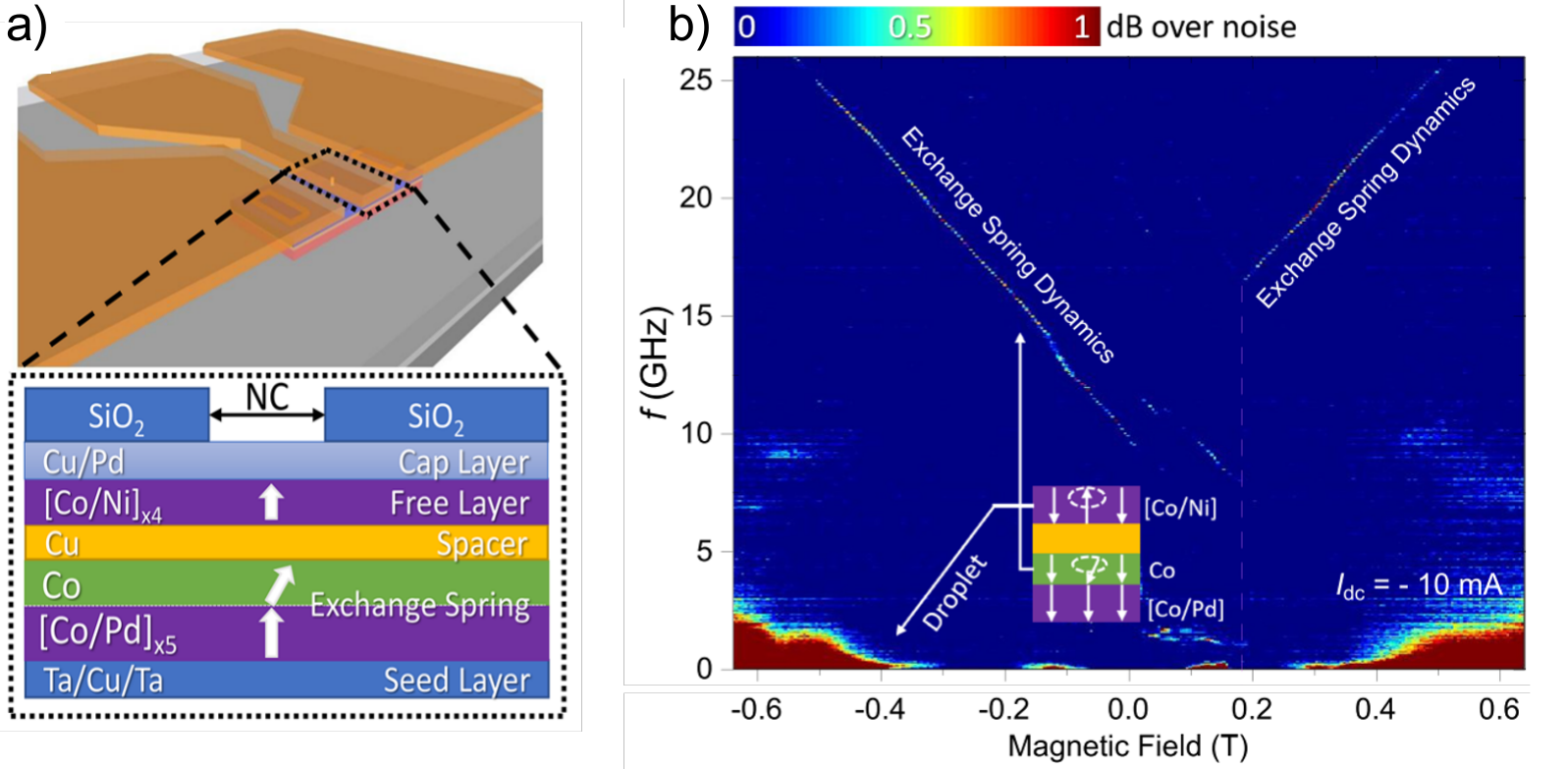}
    \caption{\textbf{a} Schematic picture of an NC-STNO device and its film stack structure, where the exchange-spring layer consists of a hard magnet [Co/Pd] with strong PMA and a soft Co layer with weak IMA. \textbf{b} Power spectral density (PSD) as a function of the OOP magnetic field. The inset is a schematic of the magnetodynamics of the two layers. \emph{Reprinted with permission from Jiang \emph{et al.}}~\cite{Jiang2023} \emph{Copyright {2023} American Chemical Society.}}
    \label{ESdroplet}
    \end{center}
\end{figure}

The droplet typically forms in magnetic materials with strong perpendicular magnetic anisotropy, where an in-plane or out-of-plane reference layer is used. Recently, Jiang~\emph{et al.}~\cite{Jiang2023} attempted to generate droplets using an exchange-spring-based RL. This reference layer comprises a thick Co layer with in-plane magnetic anisotropy (IMA) and a [Co/Pd] multilayer with strong PMA. By controlling the IMA and PMA, the exchange-coupled structure enables the realization of the tilted anisotropy illustrated in Fig.~\ref{ESdroplet}(a). Notably, the formation and dynamic behavior of the droplets were observed in the exchange-spring layer as seen in Fig.~\ref{ESdroplet}(b). These exchange-spring spin-torque nano-oscillators also function as bipolar oscillators, operating under both positive and negative currents. This significantly broadens the design possibilities and functionality of the current STNO technology for energy-efficient high-frequency spintronic and neuromorphic applications.

\subsection{Magnetic Tunnel Junction (MTJ)}
\label{sec:MD:MTJ}

Two desired properties of potential devices are tunability and a strong output signal. Magnetic tunnel junctions can be utilized to achieve both. MTJs are superior to spin valves and SHNOs in terms of power emission, since the magnetoresistance ratio can be orders of magnitude larger: 
200\%~\cite{Wang2018}, or even 600\%~\cite{Ikeda2008}. However, the high resistance tunnelling barrier makes it impossible to use the standard NC-STNO mesa design. The current will flow through the top layers to the ground, instead of getting spin-polarized during tunneling. A common method to restrain the current flow is to use the nanopillar design, but droplets have not been experimentally observed in that type of device. Instead, careful manufacturing of devices where the current is forced to pass the tunnel layer is needed.

The magnetodynamics of sombrero-shaped MTJs with PMA-free layers was studied, but 
droplets were not detected~\cite{Jiang2019}. This called for a revised device design, and droplets were later demonstrated in similar MTJs with a double free layer.~\cite{Shi2022} The new design reduced the lateral current density ($J_{\text{x}}$) by half, thereby also reducing the Zhang-Li torque, which appears to be detrimental to droplet stability. Simulations showed that the soliton also may appear in single free layer MTJs, but only in a narrow current range. In contrast, with double free layers the available current range is wide, and local inhomogeneities promote an even wider range. 


\subsection{Nanowire STNO}
\label{sec:MD:NW}

\begin{figure}
  \begin{center}
  \includegraphics[width=1\textwidth]{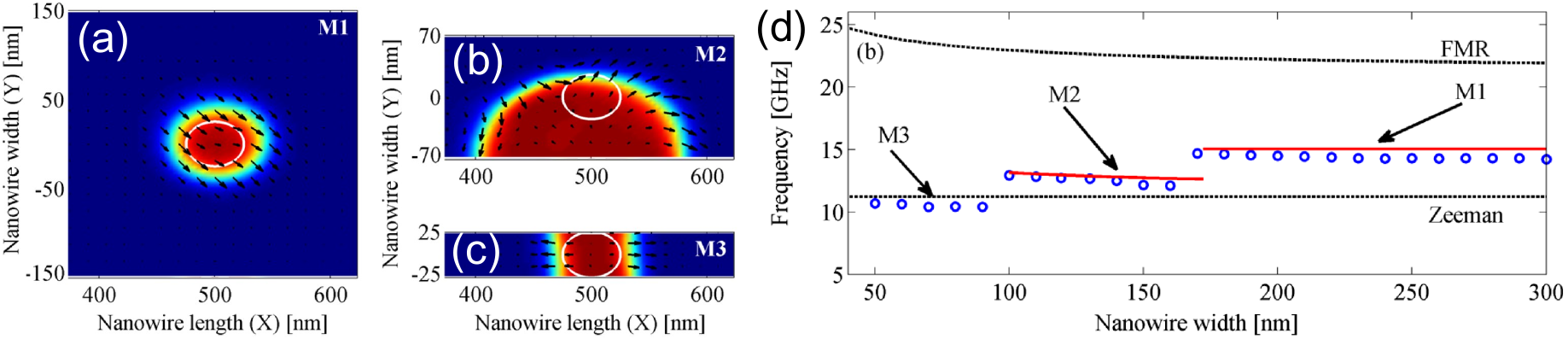}
    \caption{Characteristics of droplets in nanowires. \textbf{a} An ordinary droplet in a wide NW. \textbf{b} A half 2D-droplet attached to the NW edge. \textbf{c} A chiral quasi-1D droplet appearing in narrow NWs. \textbf{d} Droplet frequency as a function of nanowire width, together with the ferromagnetic resonance (FMR) and Zeeman frequencies. The different modes have distinct frequencies and are easily identified. The blue rings are determined by micromagnetic simulations, and the red lines are the results of analytical calculations. \emph{Reprinted with permission from Iacocca \emph{et al.}}~\cite{Iacocca2014} \emph{Copyright (2014) by the American Physical Society.}}
    \label{SMDFig1}
    \end{center}
\end{figure}

Physical confinement is an important aspect of all nanoscience. The impact of lateral boundaries on the magnetic droplet has been investigated in a publication by Iacocca~\emph{et al.}~\cite{Iacocca2014}. They used micromagnetic simulations of nanowire (NW) STNOs and discovered that the droplet envelope transforms as the NW width decreases. Three modes were identified: an ordinary droplet (M1, Fig.~\ref{SMDFig1}(a)), a half 2D-droplet attached to one NW edge (M2, Fig.~\ref{SMDFig1}(b)), and a chiral quasi-1D droplet appended to both NW edges (M3, Fig.~\ref{SMDFig1}(c)). The frequency as a function of nanowire width is given in Fig.~\ref{SMDFig1}(d), where the different modes are easily distinguished. Including temperature in the simulations gave the same results, but the frequencies were slightly shifted towards the Zeeman frequency. 

The half 2D-droplet behaves simply as an ordinary droplet in all essential means. The large footprint leads to a low frequency, but this can be described using the same analytical expression as for M1 (red lines in Fig.~\ref{SMDFig1}). Whether the droplet sticks to the upper or lower edge is a random process. The physical trigger for the mode transformation between M1 to M2 is still not fully understood.

The quasi-1D droplet exhibit a sub-Zeeman frequency, which the authors link to the topological quality of the mode. The sub-Zeeman value persists when the external field is varied, and $f(H)$ of M3 has a slope of $27$~GHz/T. Furthermore, the quasi-1D droplet is effectively described by the same theory as conservative 1D magnetic solitons in biaxial materials, with a few quantitative exceptions. The dissipative droplet only exhibits breathing (oscillating size) in non-zero applied fields, and the breathing matches the precessional frequency, not twice the precessional frequency.

Experiments on droplets in nanowires have only been reported in brief, but the results suggest a transition to the half 2D-droplet edge mode with increasing field or current~\cite{Chung2015}. A more comprehensive study is needed to confirm the findings in simulations. Nevertheless, it is important to consider the effects of physical confinement when designing spintronic droplet devices.


\subsection{Spin Hall Nano-oscillator (SHNO)}
\label{sec:MD:SHNO}

The NC-STNO has dominated droplet science thus far, but there are other mechanisms to produce spin currents/torque and overcome damping. One of them is the spin Hall effect (SHE). This mechanism occurs in materials with strong spin orbit coupling, where spin dependent scattering generates a spin current transverse to the electronic current flow~\cite{Song2020}.  Devices relying on this effect are referred to as spin Hall nano-oscillators (SHNOs)~\cite{Chen2016}. Only two layers are needed: one heavy metal with strong spin-orbit coupling, and one magnetic layer. The current flow needs to be restrained to a limited area to get substantial current densities and thereby a large spin torque. A common means to define an active area is by pattern nanoconstrictions~\cite{Demidov2014apl, Mazraati2016apl}. The current, and the concomitant Oersted field and spin torque, have a much more complex distribution in SHNOs compared to STNOs.

Divinskiy \emph{et al.} were the first to report on droplets in SHNOs, using the very term magnetic droplet soliton~\cite{Divinskiy2017}. (It can be discussed whether earlier experimental data interpreted as the dynamical skyrmion~\cite{Liu2015}, in fact corresponds to a droplet~\cite{Akerman2018}.) In Ref.~\cite{Divinskiy2017} they use a device with a nanoconstriction width of 100~nm, and utilize micromagnetic simulations to picture the droplet behavior. At the initial state, when the droplet has just nucleated, there is a striking resemblance to the nanowire setting. The droplet attaches to one of the edges and has a size of the order of the constriction radius. However, the complex effective field landscape comes into play at longer time scales. Due to in-plane torques, the droplet grows, covers the full active area, and displays intricate dynamics. Thus, the SHNO layout cannot be expected to host well-defined, stable droplets. There is an advantage, though. The current covers an extended area, albeit weakened outside the constriction. The simulations show that a droplet can escape the active area and propagate more than 1.5~$\upmu$m. Consequently, SHNOs may be a good system to experimentally study droplet propagation.
Divinskiy \emph{et al.} were the first to report on droplets in SHNOs, using the very term magnetic droplet soliton~\cite{Divinskiy2017}. (It can be discussed whether earlier experimental data interpreted as the dynamical skyrmion~\cite{Liu2015}, in fact corresponds to a droplet~\cite{Akerman2018}.) In Ref.~\cite{Divinskiy2017} they use a device with a nanoconstriction width of 100~nm, and utilize micromagnetic simulations to picture the droplet behavior. At the initial state, when the droplet has just nucleated, there is a striking resemblance with the nanowire setting. The droplet attaches to one of the edges and has a size of the order of the constriction radius. However, the complex effective field landscape comes into play at longer time scales. Due to in-plane torques, the droplet grows, covers the full active area, and displays intricate dynamics. Thus, the SHNO layout cannot be expected to host well-defined, stable droplets. There is an advantage, though. The current covers an extended area, albeit weakened outside the constriction. The simulations show that a droplet can escape the active area and propagate more than 1.5~$\upmu$m. Consequently, SHNOs may be a good system to experimentally study droplet propagation.

Another SHNO design uses nanocontacts to restrain the current flow, just as in STNOs. Unsurprisingly, droplets have been detected in these devices as well~\cite{Chen2020pra}. The magnetic layer had a crystalline easy-axis anisotropy (along $z$) slightly weaker than the demagnetization field, which resulted in a small in-plane anisotropy, and the external field was directed almost perpendicular to the film plane ($\varphi = 82\degree$) Under these conditions a mode transition occurs at cryogenic temperatures. At small currents a bullet soliton~\cite{gerhart2007prb, slavin2005prl} is formed, which transforms into the droplet soliton above a threshold current. Measurements at room temperature only reveal a single droplet mode.


\subsection{Exotic spin-orbit torque}
\label{sec:MD:ESOT}

The conventional SHNOs described in the last chapter (Sec.~\ref{sec:MD:SHNO}), only produce in-plane spin-torques by symmetry. But recent studies have found materials that generates out-of-plane spin-torque components as well, such as WTe$_{2}$~\cite{MacNeill2017}, Mn$_{3}$GaN~\cite{Nan2020}, Mn$_{3}$Sn~\cite{Kondou2021} and IrMn$_{3}$~\cite{Liu2019, Kumar2023}. Klause and Hoffman have utilized the prospect of perpendicular torques, which they call exotic spin-orbit torques, in simulations of a magnetic nanowire (NW). The benefit of this approach is that the anti-damping torque can be applied to the whole sample, without confined constrictions or nanocontacts. Yet, the sample dimensions must be small enough to allow for high current densities. They used an area of 1125$\times$250~nm$^{2}$ with a magnetic layer thickness of 5~nm.

They observed nucleation of droplets and followed the time evolution of the magnetization. First, a single droplet was formed in the middle of the nanowire. The nucleation spot was determined by the dipolar fields at zero temperatures, but became random if significant thermal fluctuations were included. A couple of nanoseconds after nucleation, the droplet had grown, and two additional droplets appeared at each side of the first one. The pair shared the same phase, while the initial droplet precessed independently. The droplets expanded with time, and yet another pair formed to the sides of the existing trio. Also this pair shared a common phase. In the next steps, droplets merged and attached to the NW boundary, forming quasi-1D droplets (see Sec.~\ref{sec:MD:NW}). When this happened, the original coherency of the spin precession was lost, and the droplet dynamics became chaotic. New small droplets emerged continuously. The continuous formation and growth of solitons could be stopped by taking advantage of that lower currents are needed to sustain, than to nucleate a droplet. By pulsing the applied current, the number of droplets could be controlled by the length and amplitude of the pulse.~\cite{Klause2022}


\section{Other Resemblant Solitons}
\label{sec:ORS} 

\subsection{AFM and FiM Drops}
\label{sec:ORS:AFM}
Alongside ferromagnetic drops, equivalent soliton solutions have also been theoretically predicted for materials with other types of magnetic ordering - antiferromagnets (AFMs) and ferrimagnets (FiMs). 

Magnetic drops in AFMs can benefit from the intrinsic ultra-fast spin dynamics inherent for these materials. Contrary to ferromagnets, uniaxial quadratic anisotropy (in the form $-K_{1} m_\text{{z}}^{2}$) does not provide necessary nonlinear coupling between magnons in AFMs. It was proposed in Refs. \cite{Kosevich1990,baryakhtar1983dynamical} to employ higher-order terms in the anisotropy energy density, i.e. $-K_{2} m_\text{{z}}^{4}$, to stabilize drops. The two anisotropy constants together with the exchange field $H_\text{{ex}}$ define the characteristic frequencies of the AFM: the antiferromagnetic resonance $\omega_{0}=\gamma \sqrt{H_{\text{ex}} (K_{1}+2K_{2})/M_\text{{s}}}$ and the maximum-opening-angle precession frequency $\omega_{\uppi/2}=\gamma \sqrt{H_{\text{ex}} K_1/M_\text{{s}}}$, which usually are in a sub-THz range for common AFM materials. The frequency of the drops is predicted to lay below the AFM resonance in the range $\sqrt{(\omega_{\text{0}}^2 + \omega_{\uppi/2}^2)/2}<\omega<\omega_0$. In contrast to ferromagnets, where the magnetization is reversed in the center of a droplet, the maximum opening angle of the  N{\'e}el vector precession is $\uppi/2$~\cite{Ovcharov2024apl}.  

The AFM drop solution requires particular signs of the anisotropy constants $K_1>0,K_2>0$, which reduces the choice of the possible materials. The required form of anisotropy is reported for hematite~\cite{Morrish1995}, where both $K_1$ and $K_2$ can be additionaly tuned by doping elements, such as Ru, Rh, Al and Ga~\cite{Hayashi2021}. These properties make hematite a promising candidate for realizing AFM droplets in experiments. A viable detection mechanism is the spin Hall magnetoresistance.~\cite{Fischer2020}

It has also been shown theoretically that conservative and STT-stabilized drops (i.e. the latter are droplets) can form in two-sublattice FiMs~\cite{zaspel2023ferrimagnetic}. It was found that the properties of a conservative FiM drop is similar to the ferromagnetic one, but with a rescaled frequency, which strongly depends on the compensation of angular momentum between the sublattices, varying from the sub-THz range in the vicinity of a compensation point to the GHz range typical for ferromagnets.

\subsection{Dynamical Skyrmion}
\label{sec:ORS:DS}

Just as droplets, dynamical skyrmions (DS) were first investigated for idealized models of magnets without dissipation~\cite{Kovalev1979, Voronov1983}. They are named after their static counterpart, the magnetic skyrmion. The common attribute is a non-zero topological charge\footnote{This value is also known as a topological index, or skyrmion number.}, which is given by $S = \frac{1}{4 \uppi} \int \textbf{m} \cdot \left( \partial_{x} \textbf{m} \times \partial_{y} \textbf{m} \right)$~\cite{Moutafis2009, Braun20121}. The topological character implicates that the state is exceedingly stable against perturbations, and this feature has motivated an enormous research interest in magnetic skyrmions~\cite{Nagaosa2013, Fert2017, Everschor-Sitte2018}. 

\begin{figure}
  \begin{center}
  \includegraphics[width=1.0\textwidth]{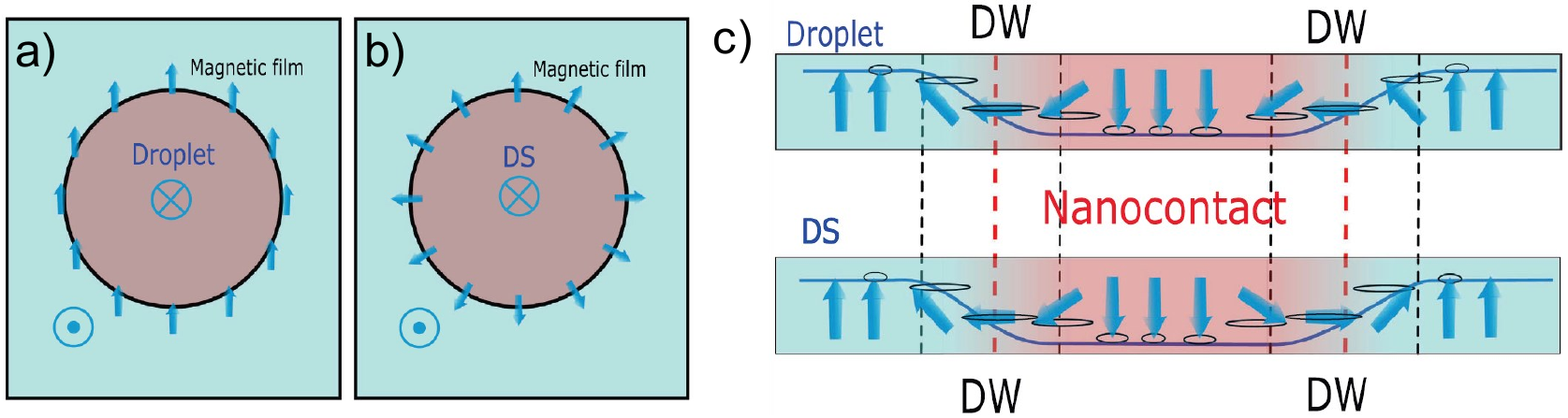}
    \caption{Illustration of the different spin textures of \textbf{a} magnetic droplets, and \textbf{b} dynamical skyrmions. \textbf{c} Cross-sections of the two solitons. \emph{From Statuto \emph{et al.}}~\cite{Statuto2018} \emph{\copyright IOP Publishing. Reproduced with permission. All rights reserved.}}
    \label{DSFig1}
    \end{center}
\end{figure}

Systems that host dissipative droplets, can also host dynamical skyrmions. There is a considerable resemblance between the two types of solitons, but the spin texture is different, giving $S = 0$ for droplets and $S = 1$ for DS. Figure~\ref{DSFig1} displays schematics of the magnetic configurations. The perimeter spins of both solitons undergo $360\degree$ precession, but while the droplet spins always keep the same phase, the dynamical skyrmion displays more intricate dynamics. The DS spins continuously alternate between all spins pointing out, clockwise alignment, all spins pointing in, and counterclockwise alignment. The dynamical soliton also exhibits breathing, i.e. the size expands and contracts.~\cite{Liu2015}

Two mechanisms can break the trivial topology of the droplet and trigger the skyrmion symmetry: i) the current-induced Oersted field~\cite{Zhou2015, Statuto2018, Zaspel2019}, and ii) the asymmetric Dzyaloshinskii–Moriya interaction (DMI)~\cite{Chen2016dmi, Song2016}. Short rise times of the applied current also favor DS formation.~\cite{Statuto2018} It is important to note that DMI is not necessary for DS formation, but in simulations this asymmetric interaction is needed in order to stabilize the soliton in zero current~\cite{Zhou2015, Carpentieri2015, Chen2016dmi, Song2016}. However, as described in Sec.~\ref{sec:M:F}, droplets freeze into static bubbles when the applied field is reduced, and they prevail when the current is switched off, pinned to magnetic inhomogeneities. Potentially, DS exhibits a similar transformation, enabling nucleation of individual nanoscale skyrmions in DMI-free materials. 

Thus far, dynamical skyrmions have only been studied by simulations. Experimental observation has been claimed~\cite{Liu2015}, but that conclusion relied on the support of simulations. Consequently, unambiguous evidence is still lacking for the existence of real DS. Experimental evidence is challenging due to the close resemblance between DS and droplets. The nucleation process is similar and only vanishingly small deviations in frequency are expected. The dynamical skyrmion differs from the droplet by being more persisting to low sustaining currents, less prone to drift and by experiencing breathing at the precession frequency.~\cite{Zhou2015, Statuto2018} Those subtle distinctions are too weak to verify DS observation beyond a reasonable doubt. 

Direct imaging of the magnetic texture is a conceivable path to distinguish the two solitons, but the necessary length scales make it a difficult task. Nevertheless, there is an experimentally accessible method to distinguish dynamical skyrmions, which was presented already in the first publication on DS~\cite{Zhou2015}. The topological soliton can injection lock to an external microwave signal via the Oersted field, while droplets cannot. This test should be straightforward and will reveal the existence of dynamical skyrmions outside the mathematical regime covered by simulations.


\subsection{More Examples}
\label{sec:ORS:ME}

Dynamical skyrmions (DS) may also be called topological droplets~\cite{Carpentieri2015}, a term which highlights the topological distinction between DS and ordinary droplets. Yet again, topology is an important quality and the use of distinctive terms benefits understanding and clear communication. Correspondingly, when any new feature of the droplet is observed, it is good to reflect on whether the discovered characteristic is inherent to “an ordinary droplet”, or if the modified state warrants a specific name. Naming of different modes usually occurs in papers presenting results of simulations, since magnetic states are easily distinguished in that kind of data. The outcome of experiments is harder to identify with a specific state, although distinct signs of mode transitions have been detected~\cite{Statuto2019, Ahlberg2022}. Based on simulations, a number of droplet states have been identified.

Unique modes appear when the standard parameters are tweaked, e.g. reducing the spatial dimensions into a nanowire leads to the formation of quasi-1D droplets~\cite{Iacocca2014}, and decreasing out-of-plane anisotropy results in spiral modes and distorted droplets~\cite{Zheng2021, Zheng2020}. Investigations of the effect of Dzyaloshinskii–Moriya interaction lead to the classification of dynamical skyrmions~\cite{Zhou2015}, instanton droplets~\cite{Carpentieri2015}, singular droplets and pseudo-normal droplets~\cite{Chen2016, Song2016}, and cigar-shaped droplets~\cite{Song2016}. The term instanton droplet refers to a mode that alternates between an ordinary zero-topology droplet and a topological dynamical skyrmion. In-coherent spin waves are emitted in the process. Singular droplets are asymmetric modes appearing at low currents, “named for the character of topological charge” (the meaning of that description is not elaborated in the original paper). The core of the pseudo-normal droplet encircles the center of the nanocontact, and perhaps this minor deviation from normal is overstressed by giving the mode a prefix. Cigar-shaped droplets carry a topological charge of $S = 1$, and exhibit a peculiar shape described by its name.   


\section{Summary and Outlook}
\label{sec:SO}

In summary, nowadays the magnetic droplet is a mature research object, and many of its mysteries have been resolved. Both experiments and simulations have provided pieces of vital knowledge needed to one day put the phenomena to use. However, simulations outnumber experiments by far, and although the interpretation of experiments needs computational aid, the validity of simulations may often be questioned. Good practice is to always include the Oe-field, which is a fundamental perturbation in a real system. Another disputed parameter is temperature. Experiments are rarely made at cryogenic temperatures~\cite{Macia2014, Lendinez2017, Statuto2019, Chen2020pra}. In contrast, most simulations ignore temperature effects with few exceptions~\cite{Iacocca2014, Puliafito2014, Zhou2015, Statuto2018}. Fortunately, any quantitative modifications of the droplet state by temperature have not been observed. The effect of thermal fluctuations seems to be limited to details, such as the droplet stability, the nucleation current, and the coherence of the dynamics---all important details in designing reliable devices, but maybe of less academic interest. Generally, simulations may be tweaked in numerous ways, but one needs experimental support to sort out what is important.

For years, the promise of using droplets in applications has been used as a motivation for further studies. Mentioned applications include magnetic random access memory (MRAM)~\cite{Okamoto2015}, magnonics~\cite{Barman_2021}, neuromorphic computing~\cite{Romera2018,Zahedinejad2020}, radio-frequency electronics~\cite{Choi2014,Sharma2021}, and Ising machines~\cite{Albertsson2021}. They all remain potential technologies, but recently a new flavor has emerged. Ferri- and antiferromagnetic materials are growing fields within spintronics and magnonics~\cite{Kim2017, Caretta2018, Jungwirth2016,Wadley2016,Baltz2018}, and we expect that droplets soon will be explored in those materials. Just as drops exist (mathematically) in both FMs and AFMs, droplets with Ne{\'e}l order will take shape in proper devices. This is an exciting next step in droplet research. Besides, the realization of propagating droplets is an appealing route towards both a more complete experimental picture and applications.	
\\
\\
\textbf{Acknowledgements} \par 
We thank B. Ivanov for valuable discussions. This work was partially supported by the Swedish Research Council (VR Grants 2016-05980 and 2017-06711) and the Knut and Alice Wallenberg Foundation.



\end{document}